\documentclass[5p]{elsarticle}

\usepackage{lineno}
\usepackage{hyperref}
\usepackage{amsmath}
\usepackage{xcolor}
\usepackage{layouts}
\usepackage{graphicx}
\usepackage{subcaption}
\usepackage{adjustbox}
\usepackage{booktabs}
\modulolinenumbers[5]

\usepackage[normalem]{ulem}
\usepackage{color}

\newcommand{\rem}[1]{}
\newcommand{\add}[1]{#1}

\biboptions{authoryear}

\makeatletter
\def\ps@pprintTitle{%
 \let\@oddhead\@empty
 \let\@evenhead\@empty
 \def\@oddfoot{\textit{\hfill\today}}%
 \let\@evenfoot\@oddfoot}
\makeatother

\begin{document}

\begin{frontmatter}

  \title{Country-wide high-resolution vegetation height mapping with
    Sentinel-2}

\author{Nico Lang, Konrad Schindler, Jan Dirk Wegner}
\address{EcoVision Lab, Photogrammetry and Remote Sensing, ETH Z\"urich}


\begin{abstract}
  Sentinel-2 multi-spectral images collected over periods of several
  months were used to estimate vegetation height for Gabon and Switzerland. A deep convolutional neural network (CNN) was trained
  to extract suitable spectral and textural features from reflectance
  images and to regress per-pixel vegetation height. In Gabon,
  reference heights for training and validation were derived from
  airborne LiDAR measurements. In Switzerland, reference heights were
  taken from an existing canopy height model derived via
  photogrammetric surface reconstruction. The resulting maps have a
  mean absolute error (MAE) of 1.7~m in Switzerland \add{and}\rem{, respectively} 4.3~m
  in Gabon \add{(a root mean square error (RMSE) of 3.4~m and 5.6~m, respectively)}, and correctly \add{estimate}\rem{reproduce} vegetation heights up to $>$50~m. They also show good qualitative agreement with existing
  vegetation height maps. Our work demonstrates that, given a moderate
  amount of reference data \add{(i.e., 2000~km$^2$ in Gabon and $\approx$5800~km$^2$ in Switzerland)}, \add{high-resolution}\rem{dense} vegetation height maps with 10~m
  ground sampling distance (GSD) can be derived at country scale
  from Sentinel-2 imagery.
\end{abstract}


\end{frontmatter}



\section{Introduction}

Vegetation height is a basic variable to characterise a forest's
structure, and is known to correlate with important biophysical
parameters like primary productivity~\citep{thomas2008}, above-ground
biomass~\citep{anderson2006} and bio-diversity~\citep{goetz2007}.
However, direct measurement of tree height does not scale to large
areas and/or high spatial resolution: in-situ observations are in
practice only feasible for a limited number of sample plots and
logging sites. Airborne light detection and ranging (LiDAR) can map
canopy height over ground densely and accurately, but the \add{financial} cost and the
limited area covered per day only allow for small regional projects
(some countries of moderate size have complete coverage, but with low
revisit times of several years between subsequent
acquisitions). Finally, space-borne LiDAR provides world-wide
coverage, but the measurements are sparse in both space and time:
distances between adjacent profiles are in the tens of kilometers, and
nearby observations have been acquired up to 6 years apart. \add{After 7 years of data collection, the point density in Gabon, for example, is only 1.26 shots per ~km$^2$ \citep{baghdadi2013viability}.} Moreover,
each measurement is averaged over a ground footprint of 70~m radius.

Hence, dense wide-area maps of canopy height are typically obtained by
regression from multi-spectral satellite images, using in-situ or
LiDAR heights as reference data to fit the regression
model~\citep{lefsky2010global,hudak2002integration}. This approach has
made it possible to produce tree height maps with ground resolutions
down to 30~m, by exploiting the Landsat archive~\citep{hansen2016}.

Here, we demonstrate country-wide mapping of canopy height with a
ground resolution of 10~m, by regression from Sentinel-2 multi-spectral
data. At such high resolutions, the spectral signature of an
individual pixel is no longer sufficient to predict tree
height. Rather, the physical phenomena underlying the monocular
prediction of tree height, like shadowing, roughness, and species
distribution give rise to reflectance patterns across neighbourhoods
of multiple pixels. It is, however, not obvious how to encode the
resulting image textures into predictive feature descriptors that
support the regression.
To sidestep this problem, we resort to deep learning. Recent progress
in computer vision and image analysis has impressively demonstrated
that very deep%
\footnote{\add{E.g., \cite{he2016deep} explore CNNs up to $>$1000
    layers.}} %
convolutional neural networks (CNNs) are able to learn
a tailored multi-level feature encoding for a given prediction task
from raw images, given a sufficient (large) amount of training data.
\add{Our experiments reveal} \rem{It turns out }that texture patterns are particularly important in areas
of high (tropical) forest, extending the sensitivity of the regressor
to heights up to $\approx$55~m.
End-to-end learning of rich contextual feature hierarchies underlies
several \rem{spectacular}successes of image and raster data analysis,
including visual recognition of objects~\citep{krizhevsky2012},
understanding human speech from
spectrograms~\citep{abdel2014convolutional} and assessment of
positions in board games like go or chess~\citep{silver2018}.

We employ a deep convolutional neural network to regress country-wide
canopy height for Gabon and Switzerland from \rem{raw}13-channel Sentinel-2
Level 2A images (corrected to bottom-of-atmosphere reflectance), using
reference values obtained from airborne LiDAR scans and
photogrammetric stereo matching as training data.
The two countries were selected because in both we have access to
reference data for training and quantitative evaluation: in
Switzerland from the national forest inventory program; in Gabon via
NASA's LVIS project.
At the same time, the two countries are very different in terms of
their geography and biomes, which supports our belief that the
proposed approach can be scaled up to global coverage.
Importantly, we also find that no long time series or multi-temporal
signatures are required. A few observations per pixel (4 to 12)
already achieve low prediction errors -- in fact, even predicting from
a single image yields fairly decent results. This means that, at the
5-day revisit cycle of Sentinel-2, we are able to obtain almost
complete coverage using only the 10 clearest images within \add{the leaf-on season (May -- September) for Switzerland or within }a period of 12 months\rem{,} \rem{even }in tropical forest regions with frequent cloud cover.

Our work is, to our knowledge, the first to demonstrate large-scale
vegetation height mapping from \add{optical} satellites at 10~m GSD. The model is
able to retrieve tree heights up to $\approx$55~m, well beyond the
saturation level of existing high-resolution canopy height maps\add{~\citep[e.g.,][]{hansen2016}}. At
the technical level, we are not aware of any other work that employs
deep CNNs for canopy height estimation \add{from optical satellite data}.

Based on the present work, the next goal is to generate a global,
wall-to-wall\rem{10~m} map of canopy height.


\section{Related work}

\begin{figure*}[t]
\centering
	\begin{subfigure}[]{0.475\textwidth}
		\centering
  		\includegraphics[width=\textwidth]{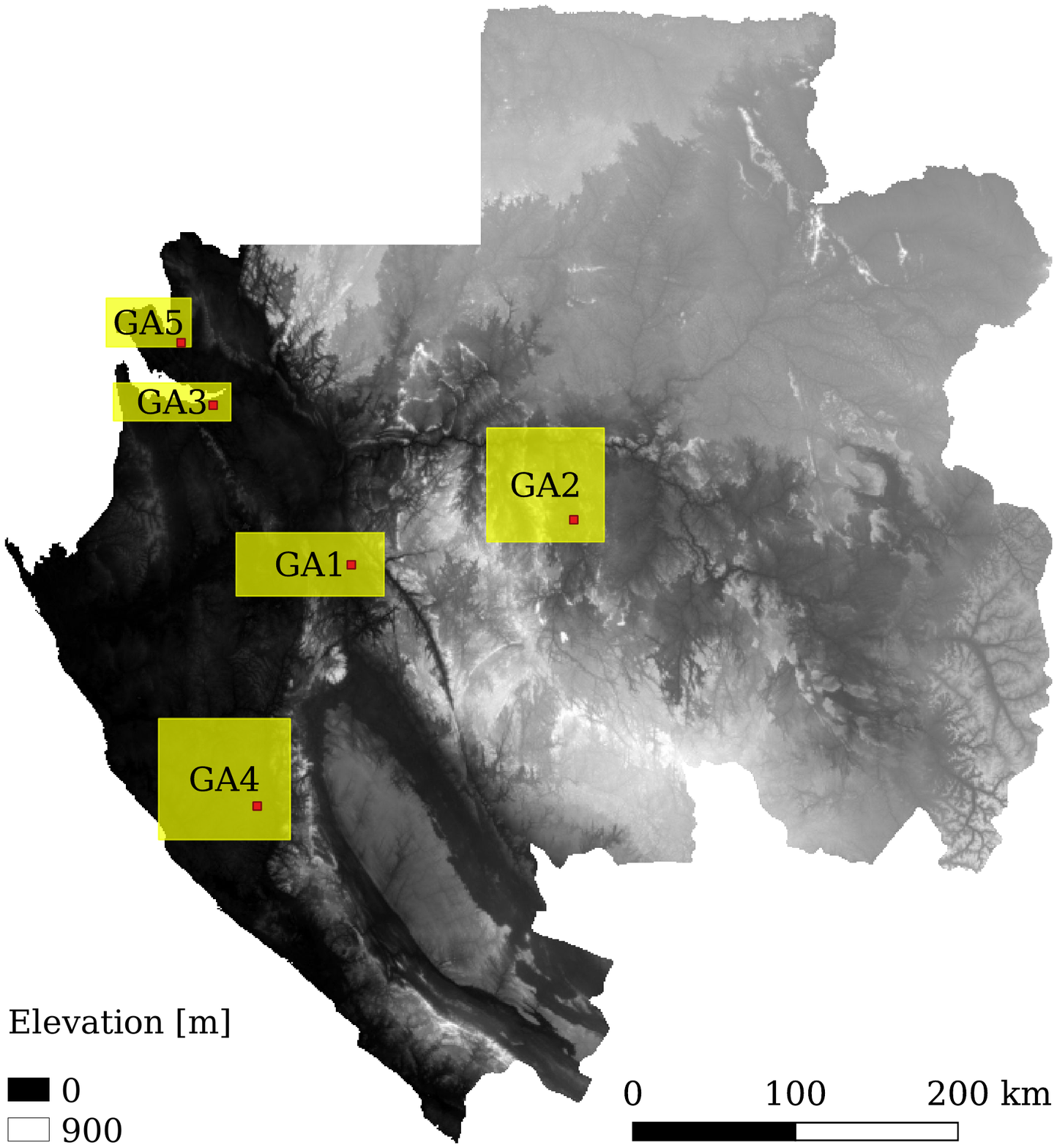}
    \end{subfigure}
    %
    %
  	\begin{subfigure}[]{0.475\textwidth}
		\centering
  		\includegraphics[width=\textwidth]{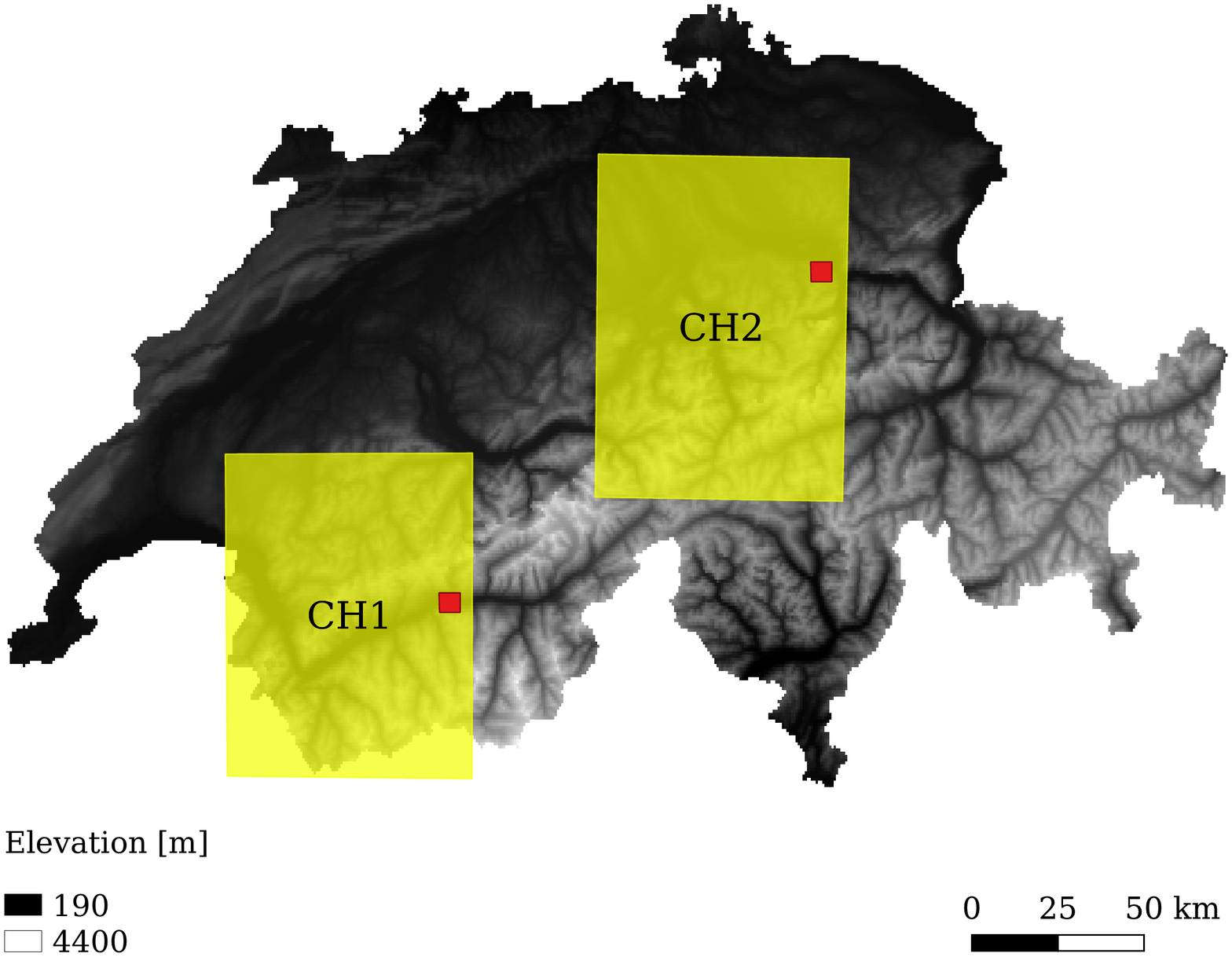}
        \end{subfigure}
        \begin{subfigure}[]{0.95\textwidth}
          \centering
            \includegraphics[width=\textwidth]{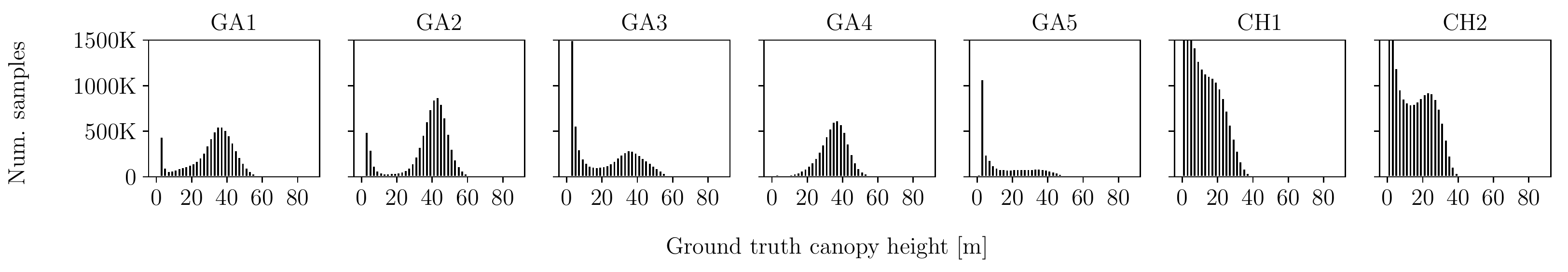}
\end{subfigure}
\caption{Overview of the ground truth regions. The red squares are the
  sub-regions shown in Figures~\ref{fig_qualitative_Gabon} and
  ~\ref{fig_qualitative_CH}. The histograms at the bottom show the
  height distribution for the seven regions.}
\label{fig_rois}
\end{figure*}

\subsection{Remote sensing of vegetation height}

The most \add{straightforward approach}\rem{natural sensor} to measure canopy height over large areas is
airborne or spaceborne LiDAR. By directly measuring range from the
sensor to both points near the tree tops and points on the ground (as
well as further ones in between), LiDAR delivers a direct and very
accurate observation of the canopy height over ground, and also makes
it possible to derive further information about vegetation
structure. That approach was developed as soon as airborne LiDAR
systems were available~\citep[e.g.,][]{naesset1997}, quickly became
popular for forest \add{monitoring}\rem{sensing} and
management~\citep[e.g.,][]{st2003tree,clark2004small}, and today is in
operational use in many countries around the world.

A limitation of airborne LiDAR is its high operating cost: due to the
limited flying height, covering large areas is time-consuming and
expensive. To scale LiDAR to wide-area (in fact, global) coverage, it
was a natural idea to deploy it from satellites, at the cost of
sparser spatial sampling. Most prominently, the Geoscience Laser
Altimeter System (GLAS) on the ICESat
mission~\citep{abshire2005geoscience} provides LiDAR profiles along
polar orbits with a footprint of 70~m, and along-track spacing of
$\approx$170~m.
A next generation spaceborne LiDAR system, the Global Ecosystem
Dynamics Investigation (GEDI) mission, has been installed on the
International Space Station. The instrument (first data release
scheduled for approximately June 2019) shall acquire 8 LiDAR profiles
along an orbit, with 25~m footprint, 60~m along-track spacing and 600~m
across-track spacing~\citep{gedi}.

Given the availability of canopy heights in some locations, and dense
global coverage by optical satellite sensors, several researchers have
attempted to fuse the two data sources into dense canopy height
maps. Technically, this amounts to regressing canopy height from
monocular (multi-spectral) images, using known tree heights as
reference data. These ``ground truth'' tree heights can be either
derived from LiDAR or by collecting enough in-situ observations.

At first sight it may appear an ill-posed task to measure canopy
height in monocular images. However, the spectral pixel signatures do
provide information about proxies like shadowing, the type of
vegetation and its density. Representative examples are for
instance~\cite{foody1996classification}, where Landsat ETM data are
used to classify rainforest into different ecological forest types,
using discriminant analysis; and~\cite{hansen2002towards}, where the
fractional tree cover is retrieved from multi-temporal signatures of
MODIS, respectively AVHRR, imagery, using a tree-based regression
algorithm.

With a similar technique,~\cite{lefsky2010global} produced a global
canopy height map, using GLAS observations as ground truth. The input
to their regression are multi-temporal optical signatures, obtained by
transforming\linebreak[4] MODIS to the \emph{brightness/greenness/wetness} space,
stacking observations from 9 consecutive months into a time series,
and compressing those to 3-dimensional signatures with principal
component analysis (PCA), independently per channel.
Remarkably, that model is able to retrieve canopy heights up to 70~m;
while its mapping resolution is \rem{of course }limited by the GSD of MODIS.

At higher spatial resolution, the main sensor used for canopy height
regression is Landsat ETM.
\cite{hudak2002integration} demonstrate kriging of vegetation height
from raw per-pixel Landsat spectra, again using LiDAR as ground truth.
The processing of \cite{ota2014estimation} is technically more similar
to the models used with moderate resolution sensors: they regress
canopy height from ``disturbance indicators'', i.e., pixel-wise
statistics over annual time series, computed over
\emph{brightness/greenness/wetness}-transformed Landsat images.
Similarly, \cite{tyukavina2015aboveground} regress canopy height from
raw Landsat data and GLAS measurements, in order to quantify forest
carbon loss over time, reaching a mean absolute error (MAE) of 5.9~m.
In follow-up work their (tree-based) regression algorithm has been
used with the same data sources, but using multi-temporal features per
Landsat channel~\citep{hansen2016}, for a 1$^\circ$ wide north-south
transect of tropical Africa. This resulted in a MAE of 2.5~m across the
entire transect (with a mean tree height $<$10~m). Whereas in regions
of high forest the heights are systematically under-estimated, with
MAEs of $\approx$5~m at 25~m tree height and $>$13~m for trees taller
than 30~m.
Consequently, the resulting maps saturate above 25~m.
We note that, interestingly, all these works base their inference only
on (in some cases multi-temporal) spectral data from a single pixel
location, although planimetric texture could potentially serve as a
further proxy signal to reveal information about the vegetation
structure, especially at high resolutions.

While our work is based on optical satellite images, tree height can
also be retrieved from RADAR. \cite{kellndorfer2014} combine Landsat
with intensity and coherence images from ALOS PALSAR-1 to map canopy
height in Chile, using again ensembles of regression trees, with
in-situ data and airborne LiDAR as ground truth. Root mean square
errors of $\approx$2~m are achieved for trees up to 30~m, whereas the
accuracy drops to $\approx$4~m if also higher trees up to 40~m are
present. In fact, their ablation studies suggest that for high trees
the retrieval is mainly supported by the optical Landsat imagery.
A variant of that method was also used to generate a world-wide map
for the tropics with GSD 30~m, based only on PALSAR-1 in
combination with GLAS ground truth~\citep{woodshole}. That map only
resolves vegetation heights up to $15$~m, a rather low cut-off for
tropical forests.
\add{While the main objective of the satellite mission TanDEM-X is the
  generation of a global digital elevation model, the data recorded by
  its Synthetic Aperture Radar (SAR) instrument was also used to
  derive vegetation heights for several forest types
  \citep{kugler2014tandem} and to map biomass in Sweden
  \citep{persson2017experiences}.}
Moreover, tree height can also be regressed from spatially explicit
maps of correlated environmental parameters (many of which are in turn
derived, at least in part, from satellite observations):
\cite{simard2011mapping} use quantities like precipitation,
temperature, elevation, tree cover etc.\ to infer canopy height, again
based on GLAS samples.

Finally, we mention that similar regression approaches have been
employed for the closely related task of (above-ground) biomass
estimation from satellite images. For instance, \cite{baccini2008} map
biomass in tropical Africa by tree-based regression from MODIS, using
in-situ observations from forest inventories and logging as ground
truth.
\cite{avitabile2012} regress forest biomass for Uganda from Landsat
imagery and land cover maps, based on ground truth field plots.
And \cite{asner2012} derive indices from Landsat imagery and SRTM
elevation data, and use them together with LiDAR ground truth to map
carbon density in Colombia.

\subsection{Forest remote sensing from Sentinel-2}

In spite of its technical superiority in terms of spatial, temporal
and spectral resolution, relatively few studies have so far used
Sentinel-2 to map forest properties, possibly because of the still
short time series.  Tree species classification has been tested for
temperate mixed forest in Central Europe~\citep{immitzer2016first} and
in Western Europe~\citep{karasiak2017mapping}.

To our knowledge, there is only one study that tests tree height
prediction from Sentinel-2~\citep{astola2019}, for boreal forest. A
two-layer perceptron is trained on a small number ($<$200) of field
plots.
They find \citep[like][discussed
  below]{chrysafis2017assessing,korhonen2017comparison} that
predictions from Sentinel-2 have slightly lower errors than those from
Landsat, in other words one does not pay a performance penalty for
going to higher spatial resolution.
There are a few studies that evaluate the retrieval of other
biophysical forest parameters, including growing stock volume for
Mediterranean forest, using regression tree
ensembles~\citep{chrysafis2017assessing}; and canopy cover as well as
leaf area index (LAI) for boreal forest, with generalised linear
models~\citep{korhonen2017comparison}.
Interestingly, the two latter studies report that prediction
performance \add{using directly the}\rem{with raw} Sentinel-2 bands was at least as good, or better,
than with vegetation indices derived from the data.

\rem{In this context it should also be mentioned that ESA's public
Sentinel-2 toolbox includes algorithms for retrieval of standard
parameters such as LAI or FAPAR, which are based on a small neural
network that was trained with simulated data to emulate the PROSAIL
radiative transfer model~\mbox{\citep{weiss2016}}}.

\subsection{Deep regression models in remote sensing}

In the last few years, deep learning with convolutional neural
networks has become a dominant technology for image analysis,
including image-level classification~\citep{krizhevsky2012},
pixel-level semantic segmentation~\citep{sermanet2014,long2015} as well
as regression of continuous variables~\citep{eigen2014}.

Deep learning has also been adopted for remote sensing, see for
example the recent overview by~\cite{zhu2017}. So far the bulk of work
applies them for classification tasks like landcover
mapping~\cite[e.g.,][]{mnih2010,chen2014deep,maggiori2017,kussul2017,marmanis2018}.

Although neural networks are a generic machine learning technology
that can, with the same machinery, solve both classification and
regression tasks, relatively few works have used them to retrieve
continuous biophysical variables or indicators.
\cite{kuwata2015} estimate crop yields from MODIS EVI and weather
maps. \cite{wang2016} retrieve sea ice concentration from Radarsat
images. \cite{srivastava2017} set up a single deep network with two
output branches to jointly solve the classification of landcover and
the retrieval of height-above-terrain (a.k.a.\ normalised digital
surface model, nDSM) from airborne G-R-NIR images. \cite{xie2016} map
a poverty indicator from downsampled \emph{Google Maps} satellite
images.  To sidestep the problem that deep learning needs large
amounts of reference data for training, they use transfer learning
from the auxiliary task of predicting night-time lights, for which
world-wide ground truth is available from NASA~\citep{noaa2014}.
Perhaps most closely related to our work,~\cite{rodriguez2018}
generate large-scale maps of tree density from Sentinel-2 images. To
obtain enough training data, they employ another deep network that
detects individual trees in \emph{Google Maps} satellite images, and downscale
the resulting maps.


\section{Data}

\subsection{Sentinel-2}

Sentinel-2 is a satellite mission within the European Space Agency's
(ESA) Copernicus program, consisting of two identical satellites
launched in 2015 and 2017, respectively, with an expected lifetime of 7.25
years. The satellites each carry a multi-spectral instrument, and
together reach a revisit time of 5 days%
\footnote{\href{https://sentinel.esa.int/web/sentinel/missions/sentinel-2}{sentinel.esa.int/web/sentinel/missions/sentinel-2}
  (2019-02-25)}. %
The sensor captures 13 spectral bands with varying spatial resolution
(10~m, 20~m, 60~m). Four bands provide 10~m ground sampling distance
(GSD), in the blue, green, red, and near infrared (NIR) regions of the
spectrum \add{\citep[further details about the band specifications can
    be found in][]{drusch2012sentinel}}. With its near and short-wave
infrared bands, Sentinel-2 is designed specifically to capture, among
others, vegetation characteristics.  The available Level 1C product
contains top-of-atmosphere reflectance values, organised in
geo-referenced 100$\times$100~km$^2$ tiles in UTM WGS84 projection.
For the present study we queried all Level 1C tiles over our regions
of interest (ROIs) in Gabon with a temporal difference to the ground
truth acquisition dates up to 9 months, and having $\leq$70\% cloud
cover. For ROIs in Switzerland we only consider\add{ed} images
captured during the 2016 leaf-on season (May -- September).  \add{By
  mixing multiple recording dates during training, the model is forced
  to acquire invariance against time-varying distractors like
  phenology or soil wetness.}

\subsection{Gabon / tropical Africa}

In February/March 2016 NASA's airborne Land, Vegetation, and Ice
Sensor (LVIS) captured full waveform LiDAR scans in five regions in
Gabon%
\footnote{\href{https://lvis.gsfc.nasa.gov/Home/index.html}{lvis.gsfc.nasa.gov/Home/index.html(2019-03-05)}}, see Figure \ref{fig_rois}. %
The measurement campaign covers mangrove forests with a maximum height
of 71.1~m in the Pongara National Park (GA3) and tropical forests with
canopy heights up to $>$85~m in the Lope National Park (GA2). With a
nominal flight altitude of 7300~m the LiDAR beams have a footprint of
18~m, with along-track and across-track spacing of $\approx$10~m.  The
LVIS data is available in two products \citep{afrisar}. The Level 1B
product contains geolocated laser return waveforms and the Level 2
product height metrics derived from them. We use the geo-referenced
canopy top and ground points from the Level 2 product to construct the
ground truth canopy height model (CHM) with 10~m GSD. In the resulting
CHM we bi-linearly interpolate missing heights that are completely
surrounded by valid canopy height values to obtain a dense ground
truth map.
We regard the heights derived from full-waveform LiDAR as ground
truth, with for our purpose negligible errors. Note, \rem{large amount
  of} \add{combining} training data \add{from several different,
  sizeable areas (in total more than 2000~km$^2$), and} recorded during
five different flights, greatly reduces the risk of systematic errors
from individual measurement campaigns.

\begin{figure}[tbh]
  \centering
  \includegraphics[width=0.9\columnwidth]{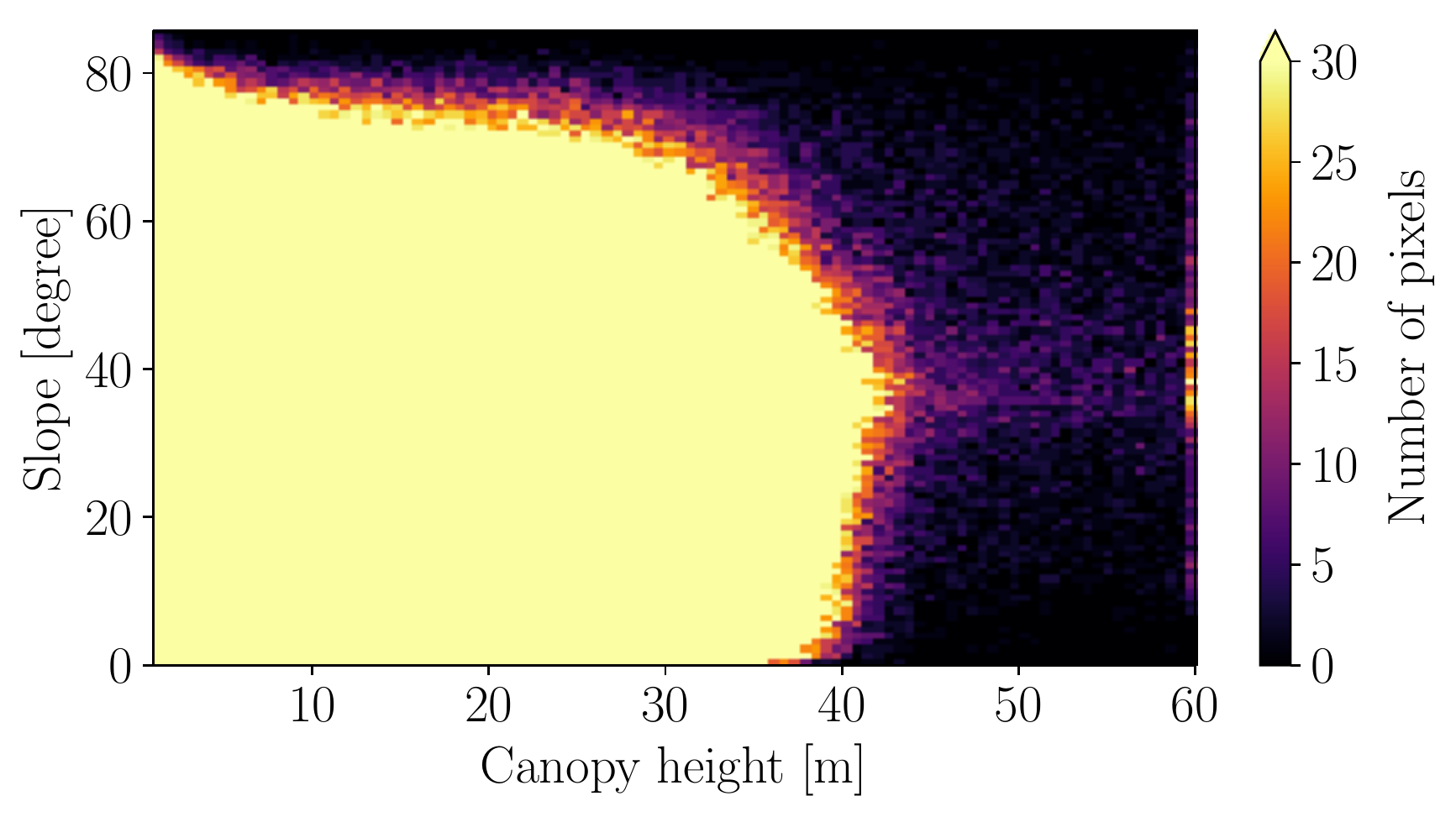}
  \caption{Switzerland ground truth: Canopy height vs.\ slope
    visualised as 2D histogram.}
\label{fig_gt_CH_heightVSslope}
\end{figure}

\subsection{Switzerland / Alps}
\label{chap_wsl_data}
The Swiss Federal Institute for Forest, Snow and Landscape
(abbreviation: WSL) created a digital surface model (DSM) by
photogrammetric stereo matching~\citep{ginzler2015countrywide}, based
on aerial imagery acquired by the Federal Office of Topography
(swisstopo). The resulting DSM was then \add{converted}\rem{reduced} to a canopy height
model by subtracting an existing digital terrain model (DTM) and
masking out the buildings. The original CHM has 1~m GSD, we reproject
it to UTM WGS84 and resample it to the 10~m GSD of Sentinel-2.
The base imagery is acquired in a continuous, cyclic scheme in a way
that all of Switzerland is covered in leaf-on conditions
(May--September) over a period of six years. We use two regions updated
in 2016, the same year for which we collect the satellite images over
Switzerland.
These regions contain parts of the Swiss Plateau, the Pre-Alps, and
the Alps, with an altitude range of 190 to 4400~m and a variety of
forest types \citep{ginzler2015countrywide}.

The accuracy of the CHM (in forest areas, after filtering out
outliers) was assessed \add{using terrestrial measurements and achieved a RMSE between 3.6~m and 5.0~m.} \rem{$\pm$3.3 to $\pm$5.5~m.}
The evaluation indicates that the canopy height errors correlate with
slope angle. In some (small) regions with steep terrain, errors in the
DSM and DTM -- probably residual mis-registration, perhaps also
slope-dependent biases -- cause grossly wrong canopy heights.
Some samples reach up to 60~m, while values $>$40~m are unrealistic for
the observed alpine forests. Figure \ref{fig_gt_CH_heightVSslope}
shows the distribution of canopy heights against the slope, where
canopy heights $>$40~m occur systematically on steep slopes
$>$30$^\circ$.  Since these samples account only for a tiny portion of
the data (0.03\% in region CH1, 0.08\% in region CH2), we simply
discard them from the test set and validate the performance only on
samples with reference height $<$40~m.
We note that (both positive and negative) biases of the ground truth
due to mis-registration can of course also occur at lower canopy
heights $<$40~m. The (small) influence of such cases is ignored.


\section{Method}

\begin{figure*}[t]
\centering
  \includegraphics[width=\textwidth]{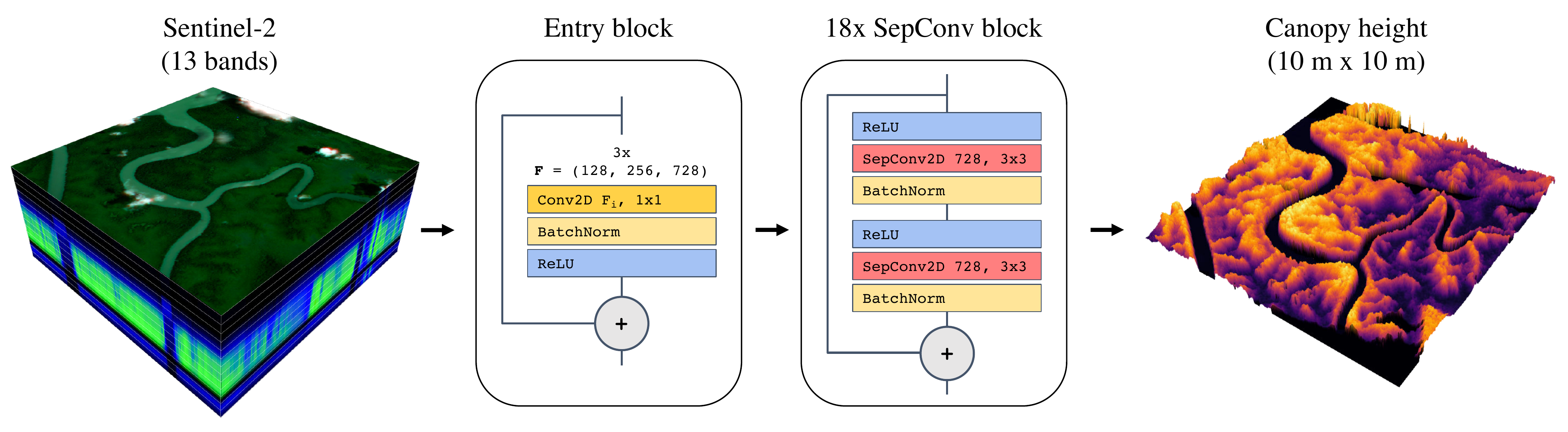}
  \caption{Method overview: A single Sentinel-2 image with 13 spectral
    bands is used as input (left) to predict canopy height at 10~m
    GSD. The entry block of our proposed CNN gradually increases the
    channel depth up to 728 channels using pointwise convolutions. The
    18 identical \textit{separable convolution} (SepConv) blocks do
    not only learn spectral features that correlate with canopy
    height, but also spatial context and texture features.}
\label{fig_methodOverview}
\end{figure*}

\begin{figure*}[t]
\centering
  \includegraphics[width=0.7\textwidth]{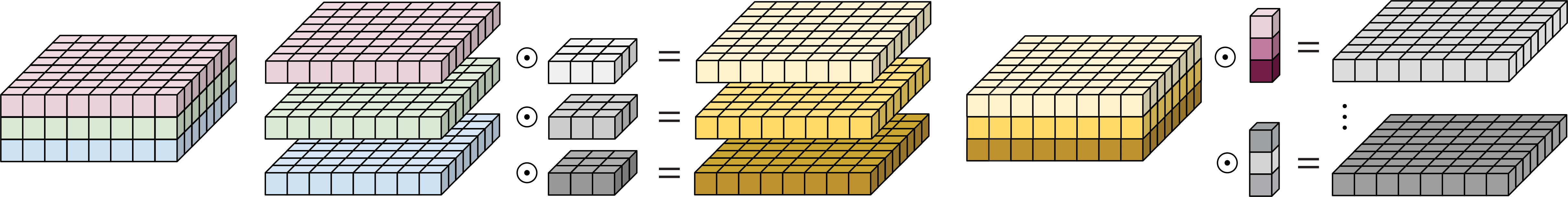}
  \caption{\add{Illustration of the depthwise separable convolution
      layer \citep[adapted from][]{sepconvEliBendersky}. First,
      separate 2D-kernels are learned for the input layers, each
      producing its own activation map. Second, these activation maps are
      combined by applying $1\times1$ kernels.}}
\label{fig_methodSepconv}
\end{figure*}

\subsection{Preprocessing}

ESA's \textit{sen2cor} toolbox provides standard algorithms to correct
atmospheric effects~\citep{step}.
As a best practice, we use this toolbox for radiometric correction
and create the Level 2A product, i.e., bottom-of-atmosphere
reflectance. By decreasing variability due to atmospheric effects, the
distribution of the image values is homogenised across different
sensing dates and geographic regions, which simplifies the regression
problem and may lead to improved generalisation.
Moreover, the Level 2A product provides also a cloud probability mask,
and a pixel-level land cover classification.  After atmospheric
correction, the lower-resolution bands (20~m and 60~m) are bi-linearly
upsampled to 10~m, to obtain a 13-channel data cube with 10~m GSD.

As a final preprocessing step we normalise each channel separately to
have mean 0 and standard deviation 1 (across the entire training set),
so as to match the requirements of the subsequent model fitting
stage. On the one hand, random initialisation then leads to neural
network weights of appropriate magnitude for all channels, which
speeds up convergence of the training. On the other hand, transforming
all feature channels to the same magnitude ensure that parameter
regularisation affects them equally.
The per-channel means and standard deviations of the training set are
stored and, at inference time, used to shift and scale the test data
in the same way.

\subsection{Deep regression network}

Convolutional neural networks (CNNs) are the state-of-the-art
technology for many image interpretation tasks, due to their ability
to learn multi-scale feature encodings with excellent predictive power
from data.
Here, we employ them for canopy height estimation from multi-spectral
Sentinel-2 images. The principle of CNNs is to extract image features
by convolving the input with a set of linear filter kernels, followed
by a point-wise non-linear activation function, to obtain so-called
\emph{activation maps}. Stacking that operation, i.e., using the
activation maps as input for another \emph{layer} of convolutions with
subsequent activation functions, yields a deep network that gradually
transforms the inputs to the desired output values via a sequence of
increasingly discriminative representations.  The parameters of the
filter kernels (also called weights) are learned directly from the
data in a supervised fashion.

Our proposed network is adapted from the Xception architecture
\citep{chollet2017xception}, see Figure \ref{fig_methodOverview}. It
consists of an \emph{entry block} followed by 18 identical
\textit{separable convolution} (SepConv) blocks.
Activation functions are always rectified linear units (ReLU) that
leave positive values unchanged and clip negative values, formally
$x_{\text{out}}=\max(0,x_{\text{in}})$.
All blocks are residual blocks, i.e., their input is also passed on
through a \emph{skip connection} that bypasses the block and is added
to the output activation map, so that the overall block learns an
additive residual function to the identity mapping. Skip connections
facilitate the learning of very deep networks by creating shortcuts
and thus preventing error gradients from vanishing before they reach
the early layers.
The entry block, inspired by~\cite{he2016deep}, consists of three
pointwise convolution layers that gradually increase the channel depth
of the data cube to 728 channels. I.e., the filter kernels are of size
$1\times 1\times d_{\text{in}}$ and together form a non-linear
per-pixel mapping to 728-dimensional spectral feature vectors.
Since the entry block changes the channel depth, the identity mapping
in the skip connection must be replaced by a single linear convolution
layer (also with learned weights) that increases the depth from 13 to
728.

The structurally identical SepConv blocks that make up the remainder
of the network each consist of two depthwise separable convolutional
(SepConv) layers and an identity residual connection that bypasses
both. A SepConv starts with the ReLU non-linearity. Then follows a
\textit{depthwise separable convolution} layer (Figure~\ref{fig_methodSepconv}). In that layer, the
overall 2D-convolution with a 3D-kernel is factored into 2D-kernels
(in our case, $3\times3$) applied to each input channel, and a 1D-kernel (linear combination) that combines the results from all
channels.
The factorisation decouples spatial and across-channel correlations,
thereby reducing the number of parameters to be learned. The number of
parameters is reduced by reusing the activation maps of the spatial
kernels ($3\times3$) as input to all subsequent point-wise kernels -- in our
case by a factor of $\approx$9.
After the separable convolution follows a batch normalisation
(BatchNorm), which renormalises the data cube for a batch (training
examples are passed through the network in small batches, see
below). The repeated re-normalisation reduces the sensitivity of the
network to the initialisation and allows for much higher learning
rates during gradient descent~\citep{ioffe2015batch}. Moreover in
conjunction with the ReLU activation it amplifies the non-linearity,
by ensuring the presence of negative activations.

After 18 SepConv blocks the final layer of our network is a pointwise
convolution that combines the 728 activation maps into a single canopy
height value per pixel.
Overall, the network has 19,604,225 trainable weights.
Since we are facing a regression problem with continuous (height)
values as targets, we learn those parameters by minimising the
$\ell_2$-loss, i.e., the mean square error over the training
examples. Furthermore, we include an optional $\ell_2$-penalty on the
parameters (``weight decay'') to regularise the fit, such that the
total loss function we minimise is
\begin{equation}
Loss=\frac{1}{N}\sum_{i=1}^{N}  \left ( f(x_i)-y_i \right )^2 + \lambda \frac{1}{W}\sum_{j=1}^{W}  \left ( w_j \right )^2
\end{equation},
with the model $f$ and its weights $w_i$ (including the constant biases
per kernel), input intensities $x_i$, ground truth canopy heights
$y_i$, and the prediction $f(x_i)$ at pixel $i$.
$N$ and $W$ denote the numbers of samples and weights, respectively.
\add{The hyperparameter $\lambda$ controls the strength of the regularization.}
We note that, in contrast to a number of CNN architectures popular in
generic computer vision, we deliberately do not down- or up-sample the
activation maps at any point. There is no \emph{max}-pooling and all
convolutions are computed with stride 1. The data cube is padded at
its borders before every 3$\times$3 convolution, to maintain the size
of the input.

\subsection{Model learning}

We use a patchwise training procedure with patches of size
$15\times15$ pixels, corresponding to $150\times150$~m$^2$ on the
ground. 
\add{These patches are sampled randomly from all valid ground truth
  locations in the training areas, consequently each patch has at
  least one valid ground truth pixel at its center. Pixels with
  missing ground truth do not contribute to the per-patch loss, such
  that they do not affect the training procedure.}
At train\add{ing} time we use\add{d} the Level 2A cloud probability mask to
exclude cloudy patches: pixels with $>$10\% cloud probability are
considered as cloudy, and any patch with $\geq10$\% cloudy pixels is
discarded. With this procedure we avoid showing the network confusing
patches with too little signal, but enable it to learn textural
features that are robust near the cloud borders.

The optimisation of the parameters in deep networks is done by
mini-batch stochastic gradient descent (SGD). To that end, one applies
the network to a (small) subset of training samples, which is called
forward pass, and from the response computes an approximation to the
gradient (partial derivative) of the loss function w.r.t.\ every
network weight.
As the CNN constitutes a nested sequence of transformations, those
derivatives can be computed with the chain rule and back-propagated
through the network, traversing it from the output to the input.
The training procedure consists of iteratively drawing batches of
training samples, back-propagating the error gradients, and updating
the weights with small steps in the negative gradient direction. The
step size is controlled by scaling the gradients with a
hyper-parameter called the learning rate.
We use a popular variant of SGD called ADAM \citep{kingma2014adam}
that adaptively adjusts the learning rate for each trainable parameter
by normalizing the global learning rate with the running average of
the gradient. This has the effect of amplifying the step size along
low gradients and attenuating it for \rem{for}high gradients. In this way,
the solver is less sensitive to the chosen base learning rate, and
there is no need to design a careful learning schedule.

For the experimental evaluation of our method the base learning rate
is set to 0.0001 and the batch size to 36 patches, dictated by memory
constraints of the GPU hardware (in our case an Nvidia GTX 1080 with 8
GB memory). I.e., during a complete training run of $\approx$50,000
iterations the network sees about 1.8 million randomly sampled
patches.

\begin{figure}[tb]
\centering
  \includegraphics[width=0.7\columnwidth]{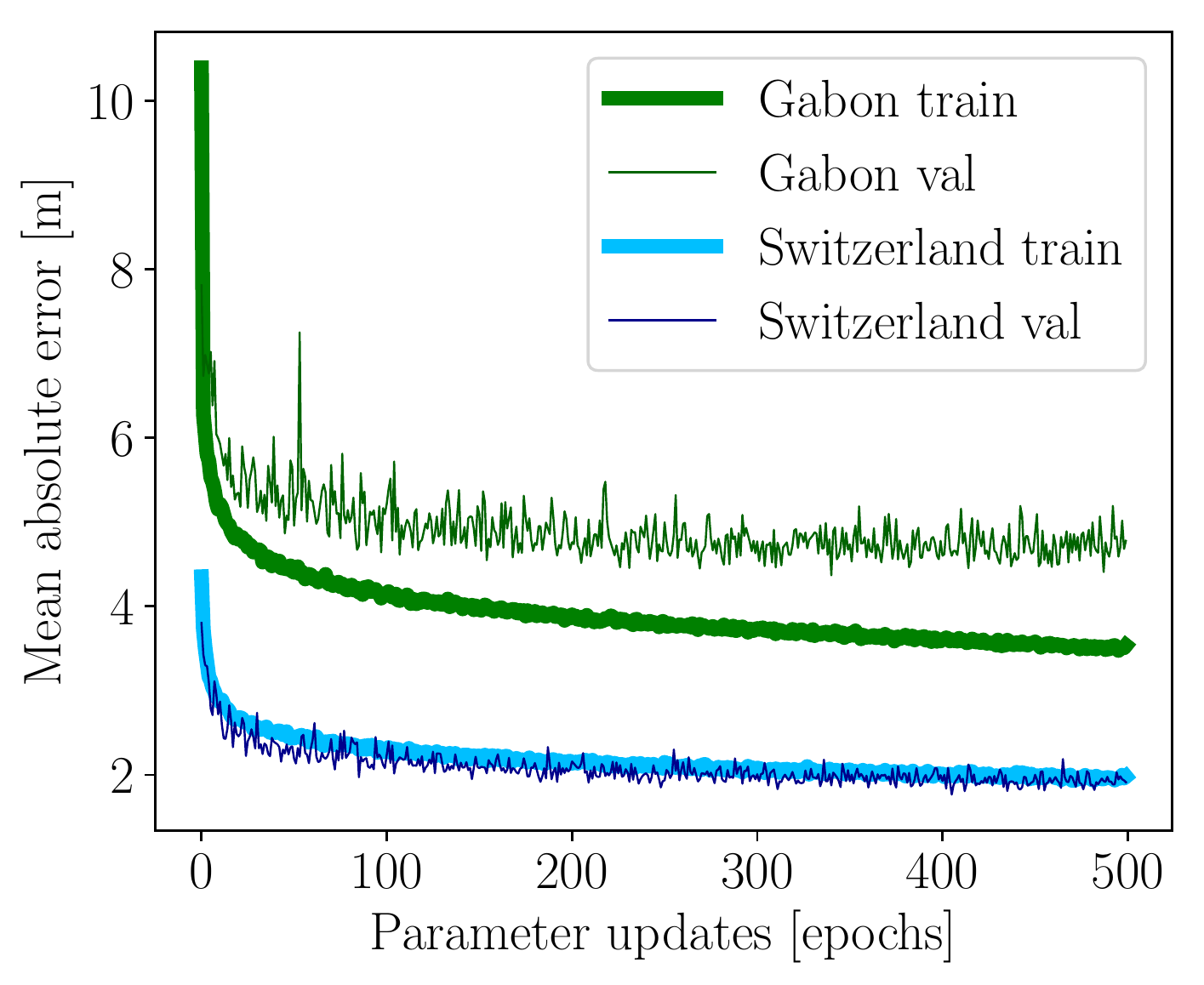}
  \caption{Loss curves for the training and validation data in Gabon
    and Switzerland. One epoch corresponds to 500 training
    iterations.}
\label{fig_training_curves}
\end{figure}

The available data is first split into two geographically separated
parts, a training set and a test set that is never seen during
training and only used to evaluate the performance of the trained
model.
The training set is, in turn, split into a training part and a
(smaller) validation part, where the latter is never used to compute
gradients for back-propagation.
The learning process is monitored by observing both the training loss
and the validation loss.
The former indicates how well the model fits the training data,
whereas the latter is \add{an estimate}\rem{a measure} of how well the model generalises to
unseen data.
Stagnation of the validation loss is a sign that the training has
converged and further iterations may cause overfitting of the training
data.
We keep training until the loss on the training set has converged,
which takes $\approx$50,000 iterations for a single training region,
and $\approx$250,000 iterations when training a single model across
all regions.
Following standard practice, we regularly (in our case every 500
iterations) calculate also the loss on the validation set. Figure
\ref{fig_training_curves} depicts the two loss curves over the
training iterations. As our final set of model parameters we pick
those that achieved the lowest validation loss.

\subsection{Evaluation}

The performance of the trained model is then evaluated on the unseen
test region, by measuring the mean absolute error (MAE),
\begin{equation}
MAE=\frac{1}{N}\sum_{i=1}^{N}\left | f(x_i)-y_i \right |\;
\end{equation}
and the root mean square error (RMSE),
\begin{equation}
RMSE=\sqrt{\frac{1}{N}\sum_{i=1}^{N} \Big(f(x_i)-y_i\Big)^2 }\;.
\end{equation}
Our proposed architecture is fully convolutional, meaning that all
layers can be applied to input images of arbitrary size, within
hardware memory limits.
To generate large-scale maps, test images are cut into tiles of
$128\times128$ pixels, with 8 pixels overlap to mitigate tiling
artifacts. We run inference on all tiles, recompose them into
a map, then mask pixels with cloud probability $>10$\%.
Moreover, water pixels according to the Level 2A land-cover
classification are excluded from the quantitative evaluation to avoid
over-optimistic results: water has a rather distinctive spectral
signature and correctly predicting 0~m canopy height over water bodies
is fairly trivial.
In Switzerland we also exclude pixels classified as snow, because snow
strongly alters the spectral signatures.%
\add{While we do not exclude the possibility that vegetation
  height could also be estimated from data with some degree of snow
  cover, this would likely require a separate, dedicated model.}


\section{Results and discussion}

\begin{figure}[tb]
\centering
  \includegraphics[width=0.8\columnwidth]{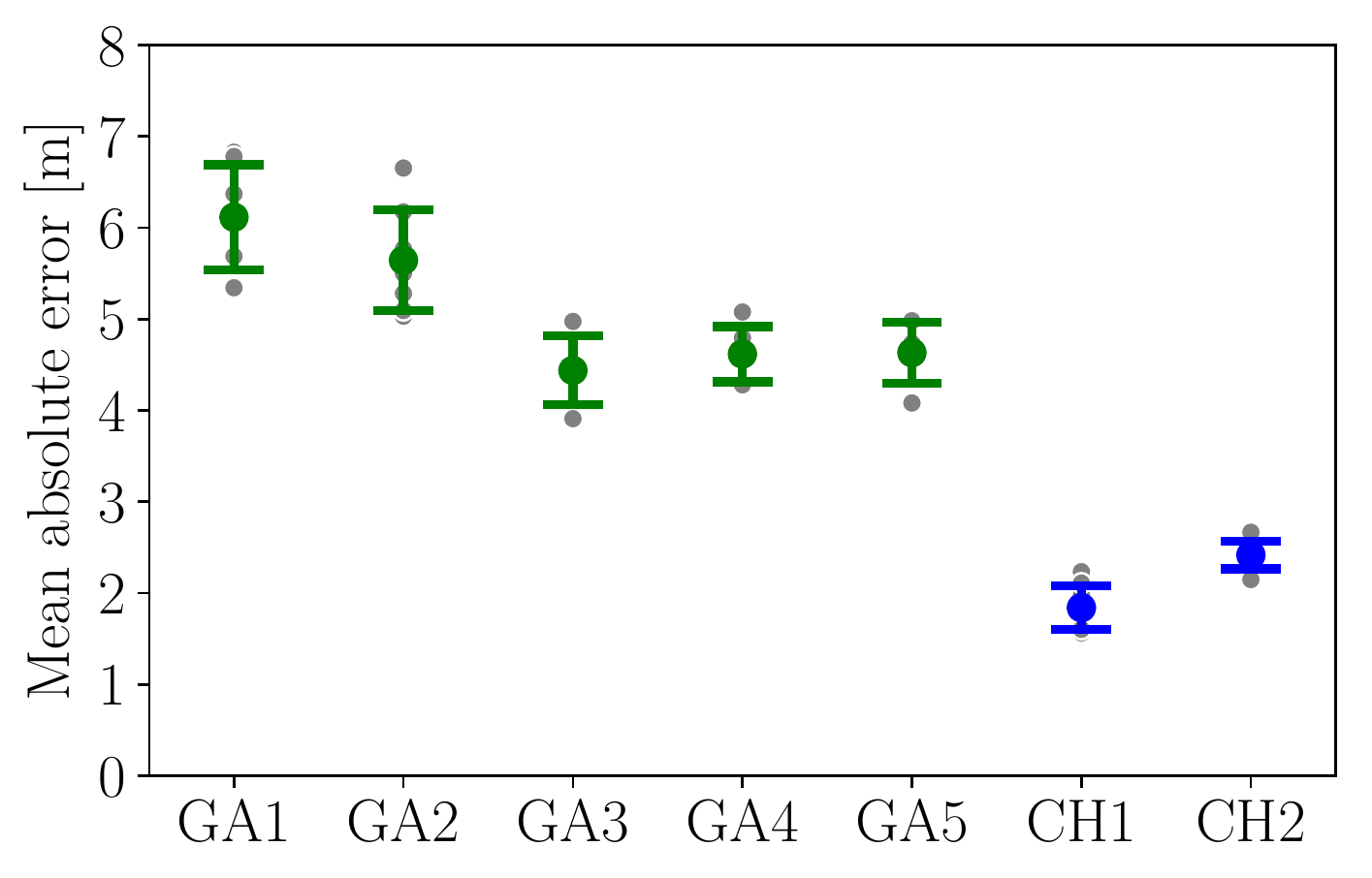}
  \caption{\add{MAEs of individual per-image predictions (grey) with the average MAE and standard deviation (error bar).}}
\label{fig_MAE_per_date}
\end{figure}

We quantitatively evaluate our approach on 7 regions in total, 5 in
Gabon (GA) and 2 in Switzerland (CH). See Figure \ref{fig_rois}. Each
region is split into spatially disjoint training, validation, and test
sets. Depending on the region, four to twelve Sentinel-2 images are
available that have overall cloud coverage $<$70\% (Table
\ref{tab_rois}). The CNN is trained on images from multiple
acquisition dates, assuming that the vegetation height did not change
significantly within the investigated time interval. This makes the
CNN more robust against radiometric variations due to remaining
atmospheric effects, illumination, vegetation activity, or water
content.
In the test set, we predict \add{high-resolution}\rem{dense}
vegetation height maps individually for each date, and merge the
resulting stack of predictions. This greatly reduces gaps due to
clouds. We have tested two simple merging strategies, choosing the
\emph{median} height and selecting for each pixel the prediction with
the lowest cloud probability (\emph{minCloud}). The median, which
benefits from the redundancy of multiple predictions and is less
affected by inaccuracies of atmospheric correction and cloud masking,
works slightly better; reaching MAE of 4.3~m for Gabon \add{and}\rem{,
  respectively} 1.7~m for Switzerland. \add{The RMSE is more sensitive
  to rare, large deviations and achieves 5.6~m and 3.4~m for Gabon and
  Switzerland, respectively.} See Table \ref{tab_crossreg}. \add{Given
  the tight correlation between the two metrics, we only show the MAE
  for further analyses.}
We point out that even the minCloud strategy, which effectively is
based on a single ``best'' spectral signature per pixel without any
temporal redundancy, achieves decent results of 4.9~m, respectively
2.0m; indicating that, if need be, one can retrieve reasonable
vegetation heights from a single cloud-free view.
\add{Furthermore, the spread between individual per-image estimates
  from different days of the year is only a fraction of the MAE: their
  standard deviation lies between 0.1~m and 0.2~m for Switzerland, and
  between 0.3~m and 0.6~m for Gabon. See Figure \ref{fig_MAE_per_date}.}
Unless explicitly stated, the reported results in the remainder of
this section are median values over all cloud-free dates.
It should be noted in this context that the multi-temporal
co-registration accuracy of Sentinel-2 is
$\approx$12m~\citep{clerc2019}, corresponding to 1.2~pixels at 10~m
GSD. Consequently, computing the median across time likely induces a
mild \emph{spatial} smoothing of the canopy height maps.

We evaluate the generalisation across sensing dates and across
geographical regions within Gabon and Switzerland,
respectively. Furthermore, an ablation study is conducted to assess
the importance of different spectral bands for canopy height
estimation. Finally, we also empirically test the influence of spatial
texture patterns, by \add{disabling}\rem{switching off} the ability of our model
to exploit spatial context.

Ablation studies are carried out on the two regions GA3 and CH2, which
are the most suitable representatives of their respective geographic
regions. Both span the whole range of canopy heights and have
a moderate cloud-cover so that generalisation across time could be
tested.
In experiments that use only a single region for training we
empirically set the regularisation of the model parameters to
$\lambda=0.001$. When training on (more and more diverse) data from
across the country, the best results were achieved without
regularisation, $\lambda=0.0$.

\begin{table}[t]
\centering
\begin{tabular}{lrrrr}
\toprule
ROI name         & Images & train [px]  & val [px] & test [px]\\ \midrule
GA1 & 6 & 4822K & 647K & 1124K  \\
GA2 & 7  & 5826K & 955K & 1575K \\
GA3 & 4 & 4393K & 1012K & 1008K \\
GA4 & 5  & 3674K & 964K & 1077K \\
GA5 & 4 & 2171K & 498K & 490K \\
CH1   & 8  & 31913K & 7499K & 9681K  \\
CH2   & 12 & 26947K & 6701K & 8461K \\
\bottomrule
\end{tabular}
\caption{Test regions with available ground truth data.}
\label{tab_rois}
\end{table}

\begin{figure}[tb]
\centering
  \includegraphics[width=\columnwidth]{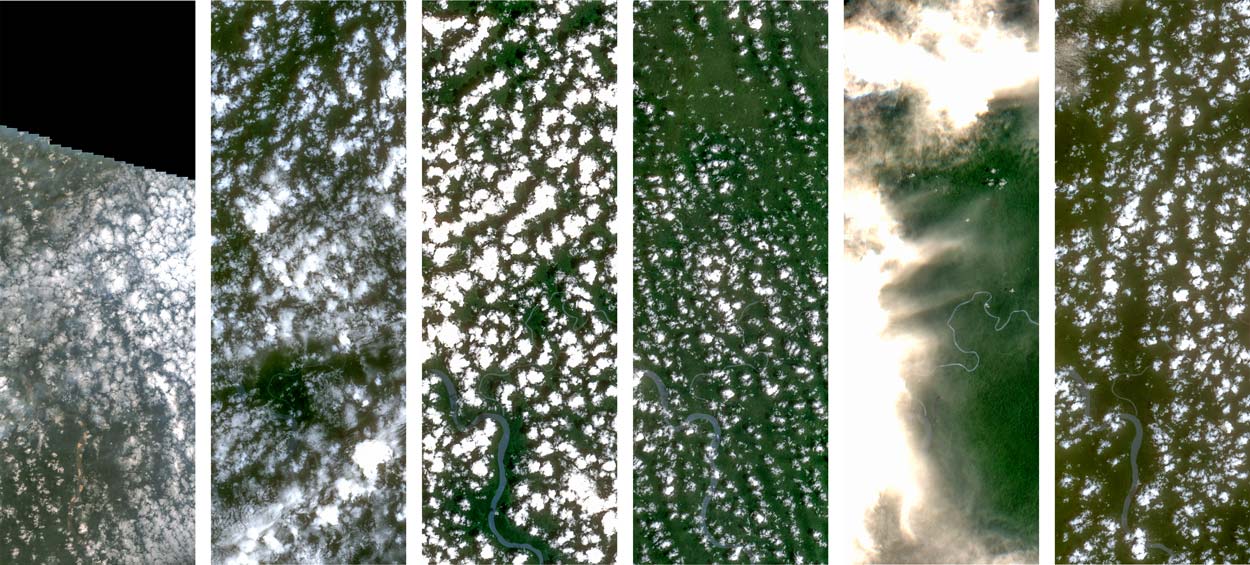}
  \caption{Example images for the test area in GA1, showing strong cloud coverage.}
\label{fig_imagesGA1}
\end{figure}

\subsection{Results for reference areas}

In Table \ref{tab_crossreg} we show the retrieval accuracy for each of
the seven regions, as well as the overall performance per country.
The CNN trained jointly on the five different training areas in Gabon
yields predictions with MAE 4.3~m across all test regions in Gabon.
The test area with the lowest error is Pongara National Park (GA3), an
area of mangrove forest, where the MAE is 3.7~m, despite being based
on only four sensing dates, due to frequent cloud cover.
In line with previous results, the predictions have higher error in
dense and high forest. In Lope National park (GA2) the MAE is
4.6~m. The worst result is 5.2~m MAE for region GA1, presumably due
to the influence of cloud shadows (Figure \ref{fig_imagesGA1}).

\begin{figure}[tb]
\centering
  \includegraphics[width=0.8\columnwidth]{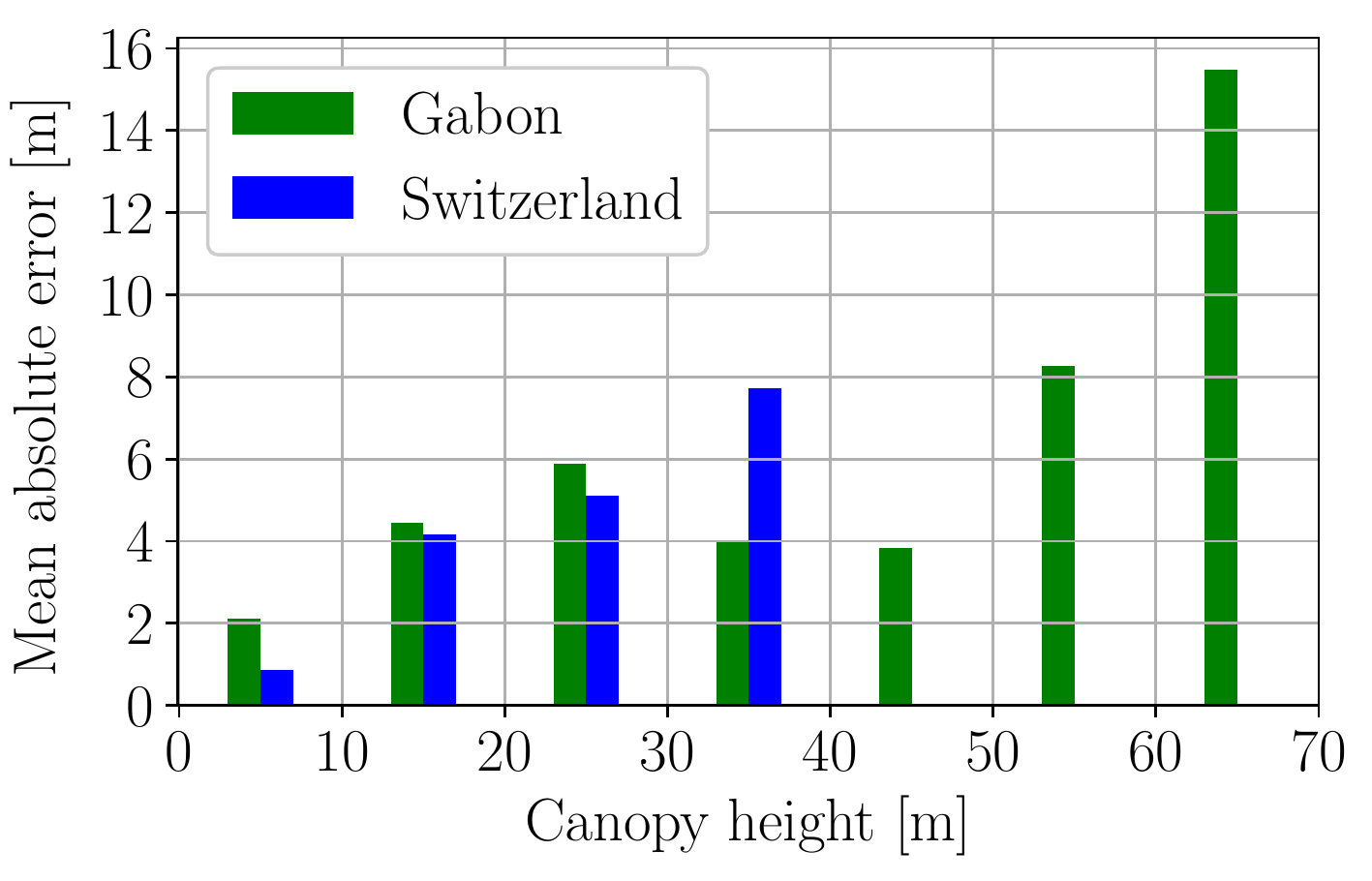}
  \caption{Mean absolute errors per 10~m canopy height intervals.}
\label{fig_bar_MAE_intervals}
\end{figure}

When fitting to both training areas in Switzerland, the predictions
across both test areas have a much lower MAE of 1.7~m.
This confirms the general trend that prediction is easier for lower
vegetation height and density. Accordingly, for CH1 with on average
lower canopy height we obtain MAE 1.5~m, whereas the higher CH2 has
MAE 2.0~m.
A more detailed view of the prediction performance in different
(ground truth) height classes is given in
Figure~\ref{fig_bar_MAE_intervals}.
Switzerland exhibits a roughly linear correlation between height and
MAE, where the relative prediction error is about 20\% across all
heights.
In Gabon the general behaviour is similar, but we observe
significantly better relative accuracy in the range of 30-55~m.

\begin{figure*}[tb]
\centering
	\begin{subfigure}[]{0.35\textwidth}
		\centering
        \includegraphics[width=\textwidth]{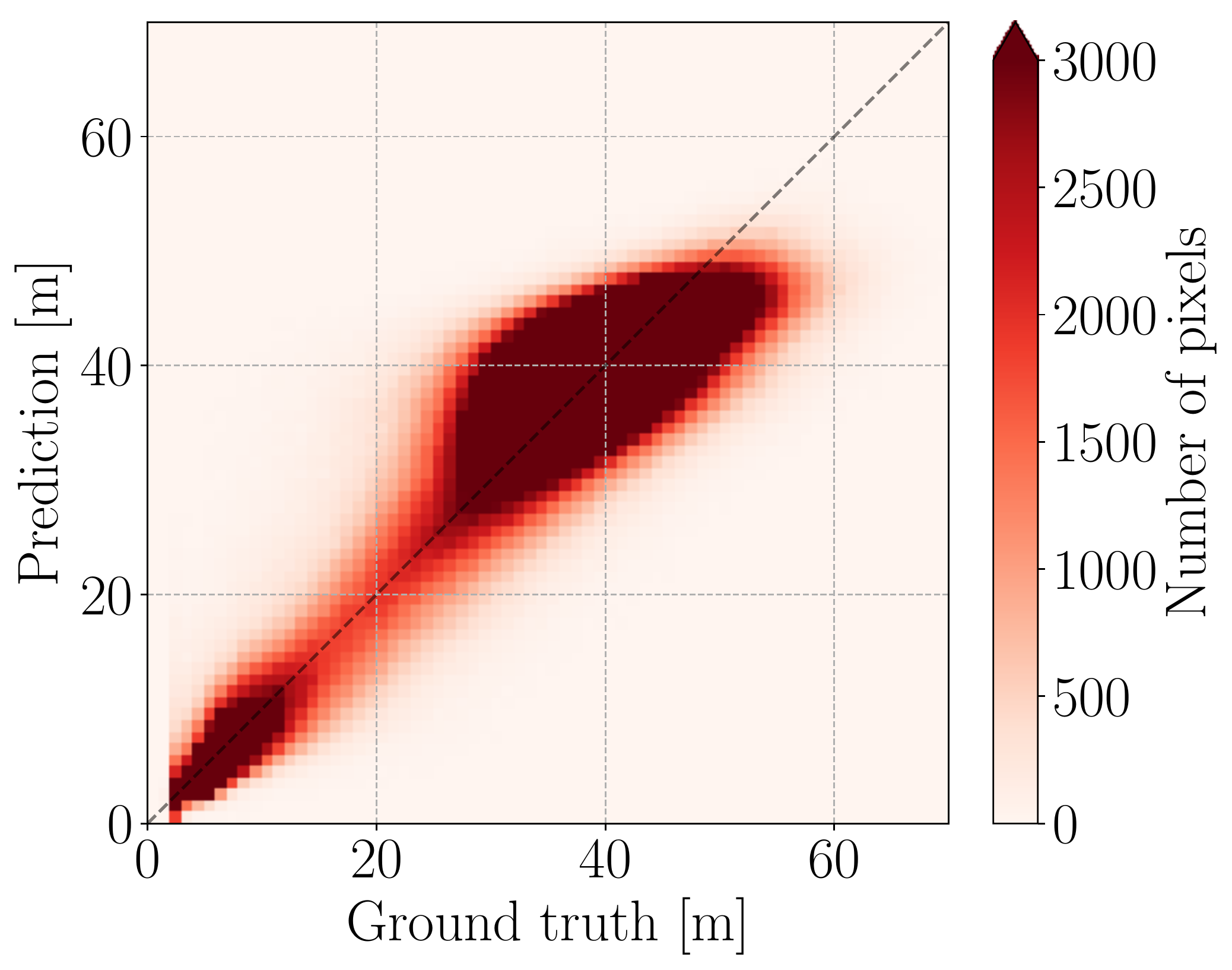}
        \caption{Gabon}
    \end{subfigure}
      \hspace{2cm}
  	\begin{subfigure}[]{0.35\textwidth}
	  	\centering
        \includegraphics[width=\textwidth]{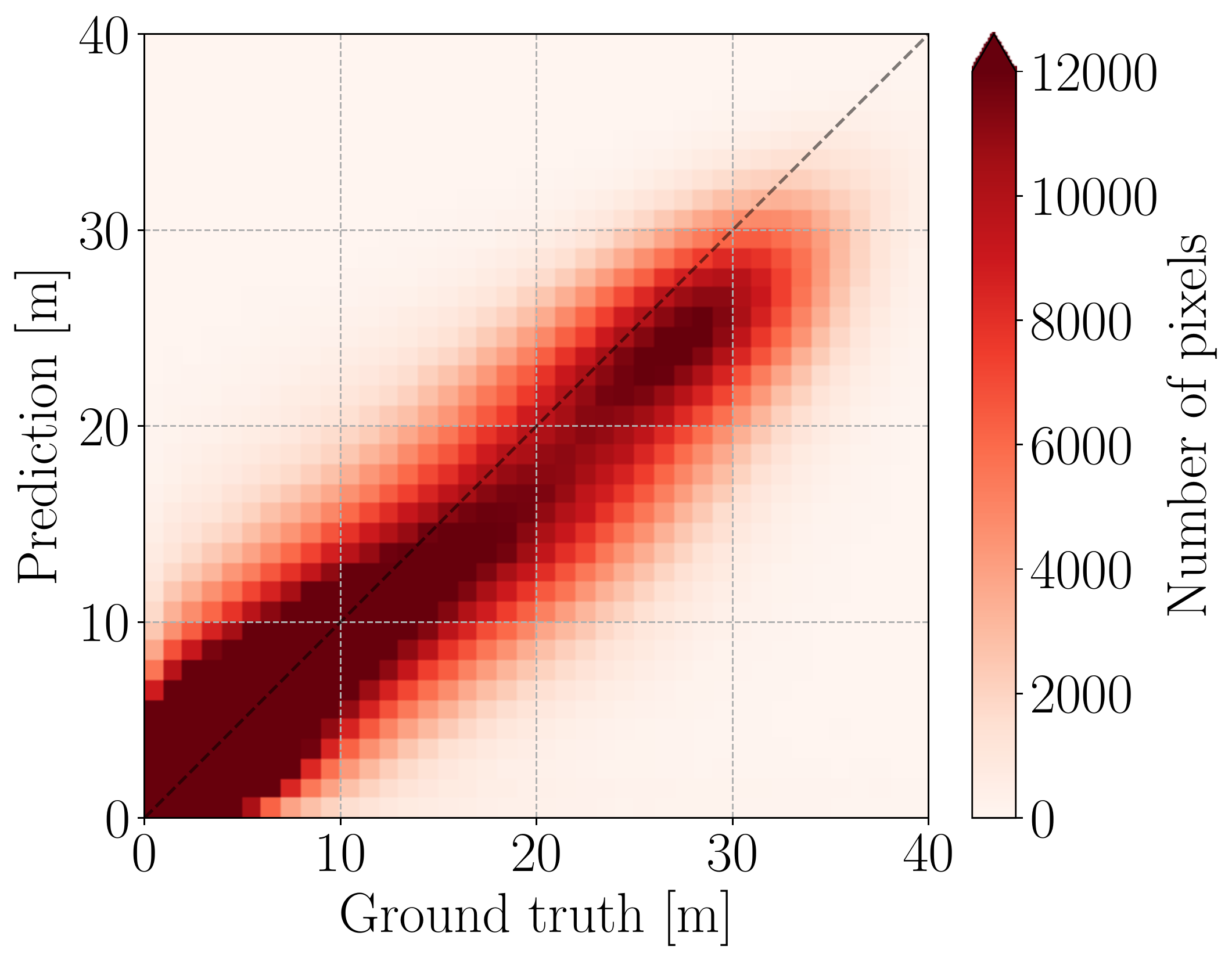}
        \caption{Switzerland}        
    \end{subfigure}
\caption{Confusion plots for Gabon and Switzerland: Ground truth
  vs.\ prediction visualised as 2D-histograms with 1~m bins.}
\label{fig_confusion_plots}
\end{figure*}

The scatter plots in Figure~\ref{fig_confusion_plots} show good
agreement between predictions and ground truth heights.
One can see that, in Gabon, there is a slight trend to overestimate
canopy heights around 30~m. A closer inspection reveals that this
happens mainly in regions densely covered by high forest, where those
30~m represent the lowest observed values.
Moreover, the model does not predict values above $\approx$55~m and
therefore underestimates very high vegetation in the range
50-70~m. This effect in high and dense forest is in line with previous
results, it appears that the canopy reflectance of rainforest
saturates at a certain height and then no longer changes \add{\citep{hansen2016, simard2011mapping}}. We find that
also the textural information is not discriminative above that height,
which is perhaps not surprising, because the species composition as
well as the shape of individual tree crowns do not vary a lot above
50~m.
In Switzerland, the predictions in general follow the ground truth
more closely. Also here, underestimation occurs for high
vegetation. Interestingly, we see an almost constant offset of
about -2.5~m for canopy height $>$15~m.
Furthermore, underestimation of low vegetation heights $<$5~m is more
frequent. In part this is probably simply a consequence of the large
proportion of samples at that height. We note, though, that a part of
the error may be due to the reference data, since photogrammetric
surface reconstruction is less reliable on low and sparse
vegetation. Also temporal changes due to agricultural plants might
play a small role (e.g., corn grows to $>$2~m). 

\begin{figure*}[h!]
\centering
  \includegraphics[width=\textwidth]{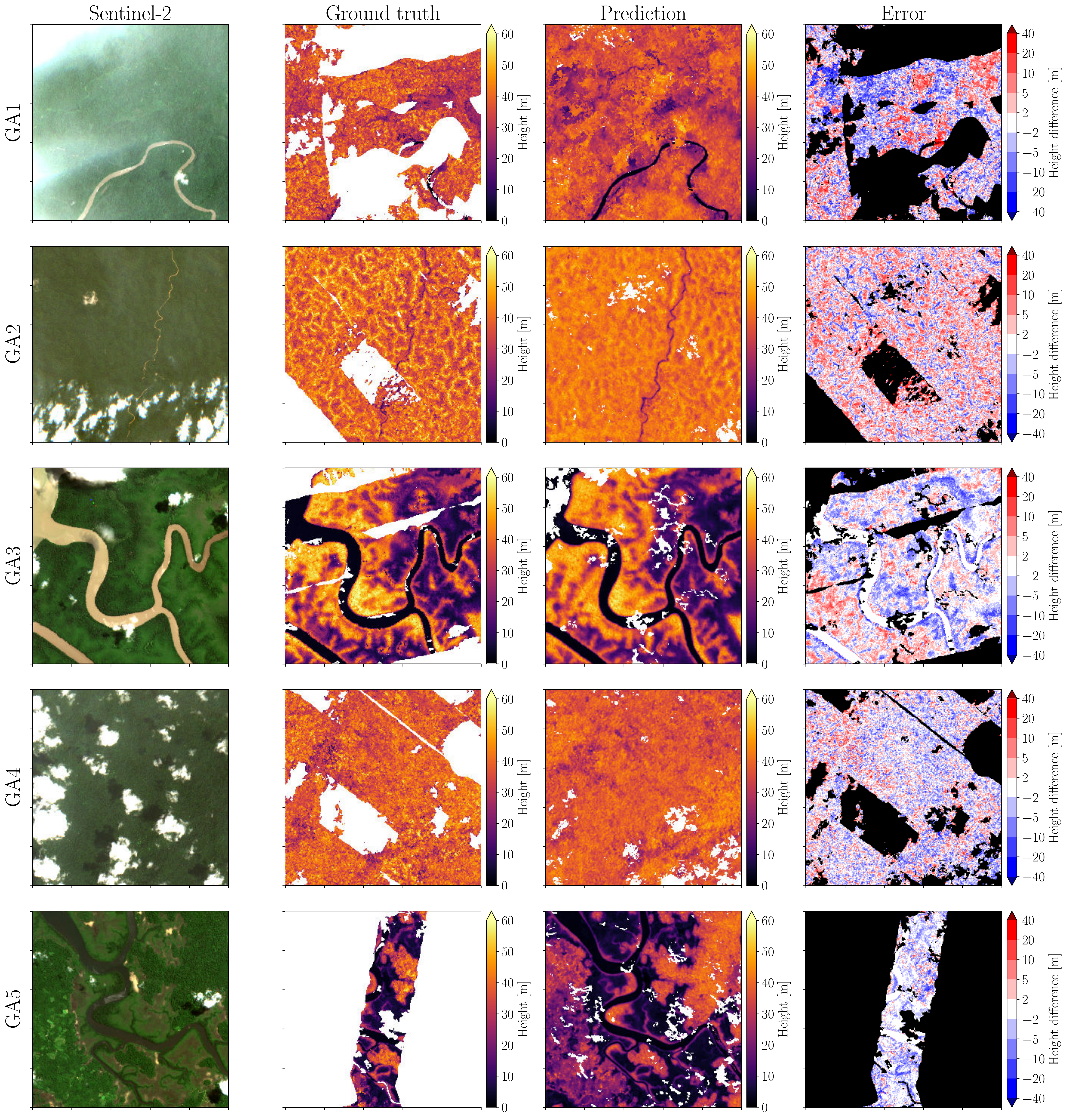}
  \caption{Qualitative results for Gabon: Rows show 5$\times$5~km$^2$
    sub-samples (10~m GSD) from the test areas of the five regions
    (GA1-5). From left to right the columns show the clearest Sentinel-2
    image, the ground truth canopy height, the prediction of our
    model, and the error (prediction minus ground truth).
    Positive errors (red) occur when the prediction is higher than the
    ground truth. \add{No data is displayed in white or black, respectively.}}
\label{fig_qualitative_Gabon}
\end{figure*}

\begin{figure*}[t]
\centering
  \includegraphics[width=\textwidth]{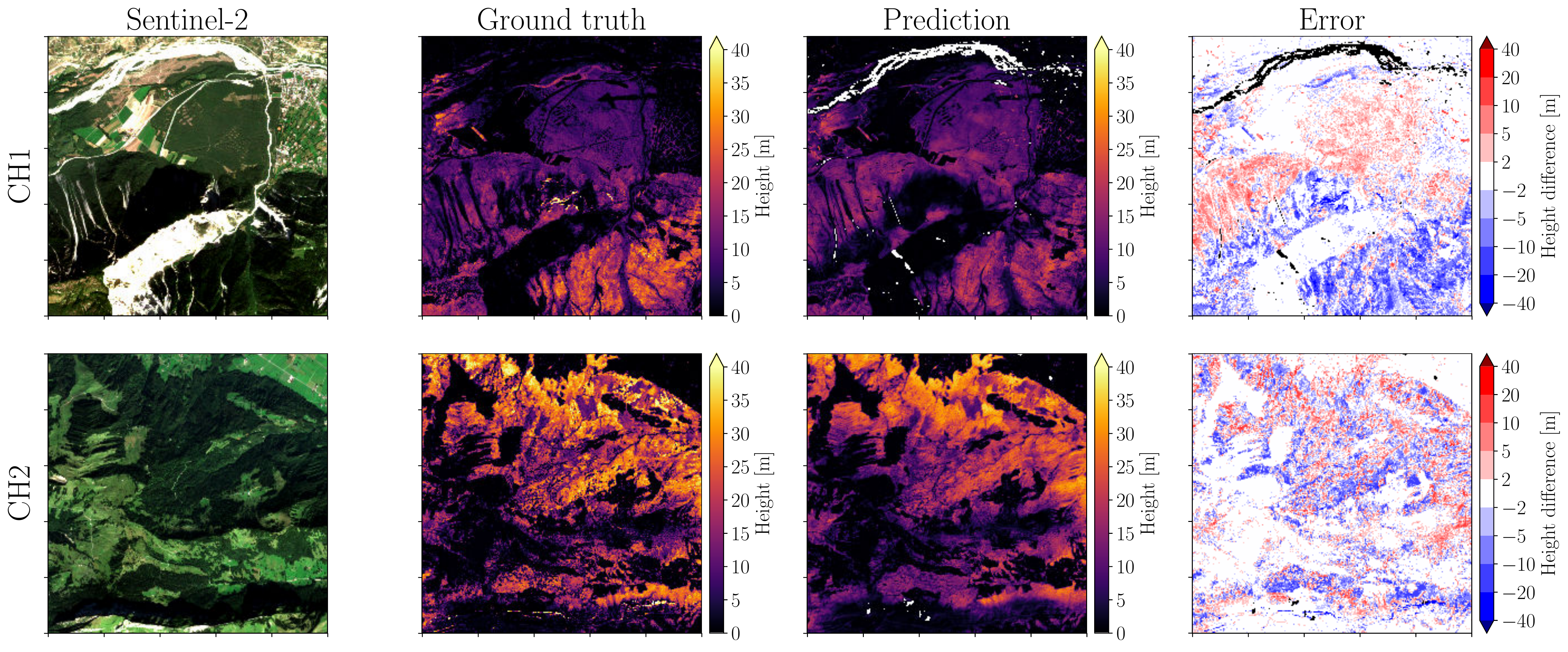}
  \caption{Qualitative results for Switzerland: Rows show 5$\times$5~km$^2$ sub-samples (10~m GSD) from the test areas of the two regions
    (CH1/2). From left to right the columns show the clearest Sentinel-2
    image, the ground truth canopy height, the prediction of our
    model, and the error (prediction minus ground truth).
    Positive errors (red) occur when the prediction is higher than the
    ground truth.  \add{No data is displayed in white or black, respectively.}}
\label{fig_qualitative_CH}
\end{figure*}

Example results are depicted in Figures \ref{fig_qualitative_Gabon}
and \ref{fig_qualitative_CH}. Each row corresponds to a different
region, showing a 5$\times$5~km$^2$ sub-sample from the test set.
The spatial distribution of the vegetation height is recovered well,
also large local variations in vegetation height are resolved
correctly (such as in region GA3).
Even fine canopy structures with extents $<$100~m are captured, but
they tend to be over-smoothed (for instance in GA2). I.e., while very
high trees $>$50~m are under-estimated (as discussed above), lower
ones near them tend to be over-estimated.
This effect seems to be a main reason for the over-estimation of
canopy heights around 30~m (see scatter plot in Figure
\ref{fig_confusion_plots}), as that is the typical height of the lower
vegetation occurring near very high stands.

\begin{table}[t]
\centering
\begin{tabular}{l c c c c c c}
\toprule
Name & \multicolumn{2}{c}{All minCloud} & \multicolumn{2}{c}{All median} & \multicolumn{2}{c}{W/O median} \\ \midrule
 & \textsc{mae} & \textsc{rmse} & \textsc{mae} & \textsc{rmse} & \textsc{mae} & \textsc{rmse} \\ \midrule
GA1 & 5.7 & 7.3 & 5.2 & 6.7 & 5.7 & 7.3 \\
GA2 & 5.4 & 6.9 & 4.6 & 5.9 & 6.9 & 8.6 \\
GA3 & 4.1 & 5.6 & 3.7 & 5.1 & 6.9 & 9.3 \\
GA4 & 4.5 & 5.8 & 4.1 & 5.3 & 4.7 & 6.0 \\
GA5 & 4.2 & 5.7 & 4.0 & 5.3 & 4.1 & 5.5 \\
CH1 & 1.6 & 3.3 & 1.5 & 3.0 & 1.9 & 3.7 \\
CH2 & 2.4 & 4.5 & 2.0 & 3.8 & 2.2 & 4.1 \\ \midrule
GA all & 4.9 & 6.5 & 4.3 & 5.6 & 6.0 & 7.9 \\
CH all & 2.0 & 3.9 & 1.7 & 3.4 & 2.1 & 3.9 \\
\bottomrule
\end{tabular}
\caption{Fusion strategies, and generalisation across different
  geographic regions in a country. The table compares MAE \add{and RMSE} (in meters)
  for: \emph{(i)} Merging multi-temporal predictions with the
  \emph{median} vs.\ the \emph{minCloud} strategy (1$^\text{st}$ and
  2$^\text{nd}$ column). \emph{(ii)} Training on training areas of
  \emph{all} regions in a country vs.\ training only on training areas
  of the 4, respectively 1, other regions (\textit{W/O}), without
  seeing data from the immediately adjacent training area
  (2$^\text{nd}$ and 3$^\text{rd}$ column).}
\label{tab_crossreg}
\end{table}

\begin{figure*}[tbh!]
\centering
  \includegraphics[width=\textwidth]{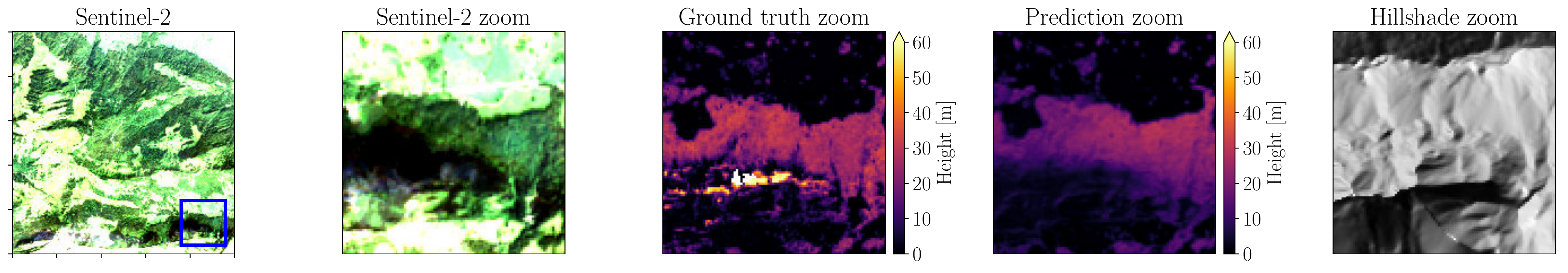}
  \caption{Illustration of DEM errors in Swiss CHM. The blue box in
    the first image marks a \add{1$\times$1~km$^2$} example area with steep
    slopes, magnified in the remaining columns.}
\label{fig_CH_GT_DTM_zoom}
\end{figure*}

In Switzerland (Figure \ref{fig_qualitative_CH}) the absolute error
for zero canopy height is $<$1~m, i.e., the model has implicitly
learned to recognise the absence of trees and bushes.
Moreover, low forest (10-20~m) tends to be underestimated, such as in
CH1.
One specific problem in Switzerland are outliers in the reference data
used as ground truth. On very steep slopes the reference data contains
narrow strips of implausible heights up to 60~m. These are likely due
to the way the CHM has been generated, by subtracting a reference
bare-earth DEM from a surface model.
Mis-registration or systematic errors on steep slopes can then cause
large differences, even in the absence of any high vegetation.
Figure~\ref{fig_CH_GT_DTM_zoom} shows an example of such a situation,
where the ground truth CHM contains values up to 60~m, whereas our
model predicts canopy height 0~m, consistent with a visual inspection
of the images.
As explained above, we therefore remove the tiny portion of samples
$>$40~m from the test set.
We did not find it necessary to remove the incorrect pixels from the
training set, as the training is apparently robust against them and
the model never predicts heights $>$40~m.
Overall, the prediction errors for Switzerland are in the same range
as the 3.3 to 5.5~m uncertainty of the canopy height model that we use
as ground truth, according to \cite{ginzler2015countrywide}. We
believe that complementary observations (e.g., airborne LiDAR) or
extensive field work would be necessary to disentangle biases of the
model from biases in the training data as well as the reference data.

In both geographic regions we observe some degree of
over-smoothing. This is a price to pay for using CNNs and modelling
texture context, as information is accumulated and mixed over larger
receptive fields.
It may be possible to mitigate the smoothing by using different loss
functions. The $\ell_2$-penalty (mean square error) used here is known
to be biased towards smooth outputs without height
discontinuities. E.g., an $\ell_1$-penalty (mean absolute error) would
instead induce a bias towards piecewise constant heights with bigger
jumps. Further investigations in this direction are left for future
work.

\subsection{Generalisation across time and space}

\begin{table*}[tb]
\centering
\begin{adjustbox}{max width=\textwidth}
\begin{tabular}{lcccccccc}
\toprule
Name & overall & 0-10 [m] & 10-20 [m] & 20-30 [m] & 30-40 [m] & 40-50 [m] & 50-60 [m] & 60-70 [m] \\ 
\midrule
GA3 & 5.7 (0.7) & 3.5 (1.4) & 5.8 (1.6) & 6.7 (1.3) & 6.5 (0.6) & 7.2 (2.0) & 9.8 (2.1) & 13.8 (2.4) \\
CH2 & 2.8 (0.3) & 1.4 (0.3) & 5.3 (0.6) & 5.8 (1.3) & 8.3 (2.0) & - & - & - \\
\bottomrule
\end{tabular}
\end{adjustbox}
\caption{MAE with cross-validation over sensing dates. Numbers in
  parenthesis are empirical standard deviations. Compare to
  2$^\text{nd}$ column of Table~\ref{tab_crossreg}.}
\label{tab_temporalCV}
\end{table*}

To analyse our model's ability to generalise to new sensing dates, we
perform leave-one-out cross-validation across different dates
(respectively, satellite passes), for the regions GA3 and CH2. All \add{retrieved}\rem{not too cloudy} acquisition dates in our considered time period except one
are used for training, the remaining one for testing.
This leads to 4-fold cross-validation for GA3, respectively 12-fold
cross-validation for CH2.
In Table~\ref{tab_temporalCV} we report the average MAE over all
folds, together with its empirical standard deviation. The baseline
performance with all dates is given in Tables~\ref{tab_ablation_Gabon}
and \ref{tab_ablation_CH}.
For the Gabon site we obtain a MAE of 5.7$\pm$0.7~m, compared to 3.7~m
when training on all dates. For the Swiss site the MAE is
2.8$\pm$0.3~m, compared to 2.6~m with all dates.
Note, the experiment is not able to separate the impact of simply
having less, and less varied, training data from the influence of not
having trained on the exact same imaging conditions. Both effects
should, according to conventional wisdom, diminish as larger training
sets with greater diversity are used.

The generalisation across geographical regions (that share reasonably
similar climate and biome) was tested by leave-one-out cross-validation
between the regions of the respective country.
I.e., we train on the training areas of all but one region in Gabon
(respectively, Switzerland) and test on the test area of the remaining
region, such that no training data from near the test area is seen
during training; resulting in 5-fold, respectively 2-fold
cross-validation.
The outcomes of that test are summarised in Table~\ref{tab_crossreg}.
As expected from Tobler's First Law of Geography\footnote{Tobler's First Law of Geography: "Everything is related to everything else, but near things are more related than distant things."}, the performance drops a bit for all
regions, as the domain gap between training and test data increases
with growing distance.
The largest drops are observed for Lope National Park (GA2) from 4.6~m
to 6.9~m and for Pongara National Park (GA3) from 3.7~m to 6.9~m.
While again the larger MAE is a combination of seeing less data and
seeing less representative data, in these cases the latter effect is
very likely the dominant cause, as these two regions exhibit rather
unique vegetation characteristics not present in the remaining
regions. GA2, in particular, has very high vegetation with average
height $>$40~m in the test area, a situation which the model is not
exposed to when training on the remaining regions.
The assertion that each of those two regions has a significant domain
gap to all others is supported by the fact that for the other three
regions in Gabon performance drops only by 0.1 to 0.6~m.

For Switzerland the performance when training on one region and
testing on the other only decreases by 0.4~m, respectively 0.2~m.
Since we only have data from two distinct regions of similar size, but
fairly similar vegetation type, that drop is probably mostly caused
by the lower amount of training data (roughly half).


\subsection{Ablation study: spectral bands}

To gain some insight into the importance of different spectral bands of
Sentinel-2 for vegetation height mapping, we train and test with
several different band combinations.
The following subsets of bands have been evaluated, by simply changing
the spectral channel depth of the input and retraining, leaving all
other specifications of the CNN unchanged. \emph{RGB}: the visible
bands with 10~m GSD (B02, B03, B04); \emph{N}: only the 10~m NIR band
(B08); \emph{RGBN}: all four 10~m bands; \emph{woRGBN}: all 20~m and
60~m bands (B01,B05-07,B08a-12); \emph{ALL}: all 13 bands.
In Tables \ref{tab_ablation_Gabon} and \ref{tab_ablation_CH} the
results are shown for the regions GA3 and CH2. For both regions the
performance of \textit{ALL} and \textit{RGBN} are similar, indicating
that the high-resolution bands carry most of the relevant information,
whereas the other nine bands (upsampled to 10~m GSD) contribute only
little.
A tentative common pattern is that towards the top of the height range
the 20~m and 60~m bands seem to \add{rather deteriorate than improve}\rem{hurt rather than help} the
regression. This would be consistent with the observation that for
high trees texture features are more beneficial (see below). But
further investigations with more geographic diversity are needed to
confirm that trend.
In Gabon, only \textit{RGBN} also achieves markedly lower MAE for very
low vegetation. Further research is needed to understand why, one may
speculate that it has to do with the implicit spatial smoothing of the
spectral information when upsampling low-resolution bands to 10~m.
All other band selections perform significantly worse in Gabon.

In Switzerland one even obtains a slight improvement when using only
\textit{RGBN} instead of \textit{ALL} channels -- again it is not
statistically meaningful to assess the significance of that result
based on one test region, and further tests are needed.
Contrary to Gabon, even using only the visible \textit{RGB} works
rather well.
Whereas the perhaps clearest message of the study is that using only
near-infrared is not sufficient.

\begin{table*}[tb]
\centering
\begin{adjustbox}{max width=\textwidth}
\begin{tabular}{lcccccccc}
\toprule
Name & overall & 0-10 [m] & 10-20 [m] & 20-30 [m] & 30-40 [m] & 40-50 [m] & 50-60 [m] & 60-70 [m] \\ \midrule
ALL & \textbf{3.7} & 2.6 & 3.7 & \textbf{4.4} & \textbf{3.8} & \textbf{4.5} & 5.9 & 7.5 \\
RGB & 5.0 & 2.7 & 5.4 & 6.9 & 5.9 & 5.1 & 7.2 & 10.3 \\
N & 6.0 & 2.7 & 5.4 & 8.1 & 7.6 & 8.7 & 8.3 & 8.6 \\
RGBN & 3.8 & \textbf{1.8} & \textbf{3.4} & 5.1 & 4.8 & 5.3 & \textbf{5.8} & \textbf{6.9} \\
woRGBN & 4.8 & 2.0 & 4.7 & 5.8 & 5.2 & 7.6 & 11.2 & 14.5 \\
\midrule
ALL 1$\times$1 & 6.0 & 1.8 & 5.1 & 7.7 & 6.6 & 11.3 & 17.1 & 22.0 \\
\bottomrule
\end{tabular}
\end{adjustbox}
\caption{Ablation study Gabon (GA3): MAE for various band selections, and
  for strictly \add{pixel-wise} spectral features (ALL 1$\times$1).}
\label{tab_ablation_Gabon}
\end{table*}

\begin{table*}[tb]
\centering
\begin{adjustbox}{max width=\textwidth}
\begin{tabular}{lcccccccc}
\toprule
Name & overall & 0-10 [m] & 10-20 [m] & 20-30 [m] & 30-40 [m] \\ \hline
ALL & 2.6 & \textbf{0.9} & \textbf{4.4} & 6.0 & 8.6 \\
RGB & 2.5 & 1.3 & 4.4 & 5.1 & 8.8 \\
N & 3.3 & 2.0 & 6.0 & 6.0 & 8.3 \\
RGBN & \textbf{2.2} & 1.1 & 4.5 & \textbf{4.9} & \textbf{7.1} \\
woRGBN & 2.7 & 1.2 & 4.9 & 6.3 & 9.9 \\
\midrule
ALL 1$\times$1 & 3.6 & 2.2 & 4.4 & 6.8 & 12.8 \\
\bottomrule
\end{tabular}
\end{adjustbox}
\caption{Ablation study Switzerland (CH2): MAE for various band selections, and
  for strictly \add{pixel-wise} spectral features (ALL 1$\times$1).}
\label{tab_ablation_CH}
\end{table*}

\begin{figure*}[t]
\centering
	\begin{subfigure}[]{0.2\textwidth}
		\centering
  		\includegraphics[width=\textwidth]{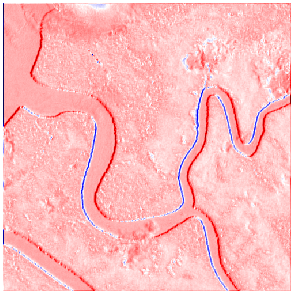}
  		\caption{}
    \end{subfigure}
    \hspace{0.5cm}
	\begin{subfigure}[]{0.2\textwidth}
		\centering
  		\includegraphics[width=\textwidth]{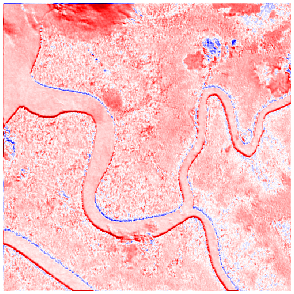}
  		\caption{}
    \end{subfigure}
    \hspace{0.5cm}
    \begin{subfigure}[]{0.2\textwidth}
		\centering
  		\includegraphics[width=\textwidth]{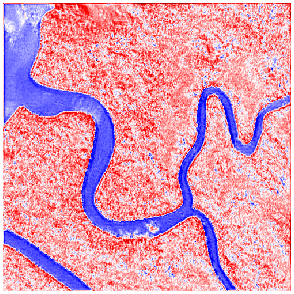}
  		\caption{}
    \end{subfigure}
    \hspace{0.5cm}
    \begin{subfigure}[]{0.2\textwidth}
		\centering
  		\includegraphics[width=\textwidth]{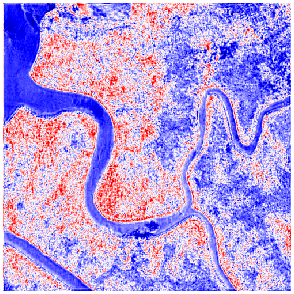}
  		\caption{}
    \end{subfigure}

\caption{Activation maps of selected $3\times3$ filter kernels our
  model learns. Left corresponds to shallower, right to deeper network
  layers. White means low activation, red/blue are strong
  positive/negative responses. One can see sensitivity to vertical and
  horizontal edges (a,b), water (c) and high vegetation (d).}
\label{fig_activations}
\end{figure*}

\subsection{Influence of texture features}

One reason for using a CNN as regressor has been our \add{assumption}\rem{conviction} that
at resolutions as high as the 10~m of Sentinel-2, where single trees
can be as big as the pixel footprint, textural features \add{may} play an
important role.
By stacking many convolutional layers with 3$\times$3 filter kernels,
our network learns to pick up texture patterns in a larger receptive
field centred at each pixel, if they correlate with vegetation height.
To check whether this is indeed necessary, we restrict the network to
look only at the spectral distribution of each individual pixel, by
setting the spatial size of all convolution kernels to $1\times1$.
This prevents it from seeing any texture or spatial context.
Otherwise the CNN architecture (number of layers, channel depth per
layer) is the same, so that the restricted network derives its
prediction at a given pixel only from that pixel's spectral
intensities, but still allows for the same degree of non-linearity in
the mapping.
The results are also reported in Tables \ref{tab_ablation_Gabon} and
\ref{tab_ablation_CH}, denoted as \textit{ALL 1$\times$1}. For GA3 the
MAE grows from 3.7~m to 6.0~m, for CH2 it increases from 2.6~m to
3.6~m. The biggest differences occur in the height range of 40-60~m.
This demonstrates the benefit of including texture features when
dealing with high-resolution images, especially in areas of very high
vegetation. Previous work dealing with vegetation height and
structure~\citep[e.g.,][]{ota2014estimation,tyukavina2015aboveground,hansen2016}
has mostly ignored texture.
There may indeed not be as much benefit in looking beyond per-pixel
signatures when one deals with lower sensor resolutions like the 30~~m
of Landsat, or even less.

In Figure \ref{fig_activations} we illustrate a few of the learned
texture features that have a clear interpretation. We show the
activation maps after applying the respective convolution
kernel. Filters in early layers extract low-level primitives such as
vertical and horizontal contrast edges. Deeper layers combine those
lower-level activations over larger receptive fields into more
task-specific features that fire, for instance, on high vegetation or
on water.

An open question in this context is whether spatial textures should be
used in conjunction with longer time series and more sophisticated
multi-temporal representations than our simple median, to further
improve the predictions.
The fact that we can get very reasonable predictions even from single
images could also indicate that texture might, to some degree, be able
to compensate for temporal redundancy and vice versa.

\subsection{Large-scale canopy height maps}

\begin{figure*}[t]
\centering
	\begin{subfigure}[]{0.49\textwidth}
		\centering
  		\includegraphics[width=\textwidth]{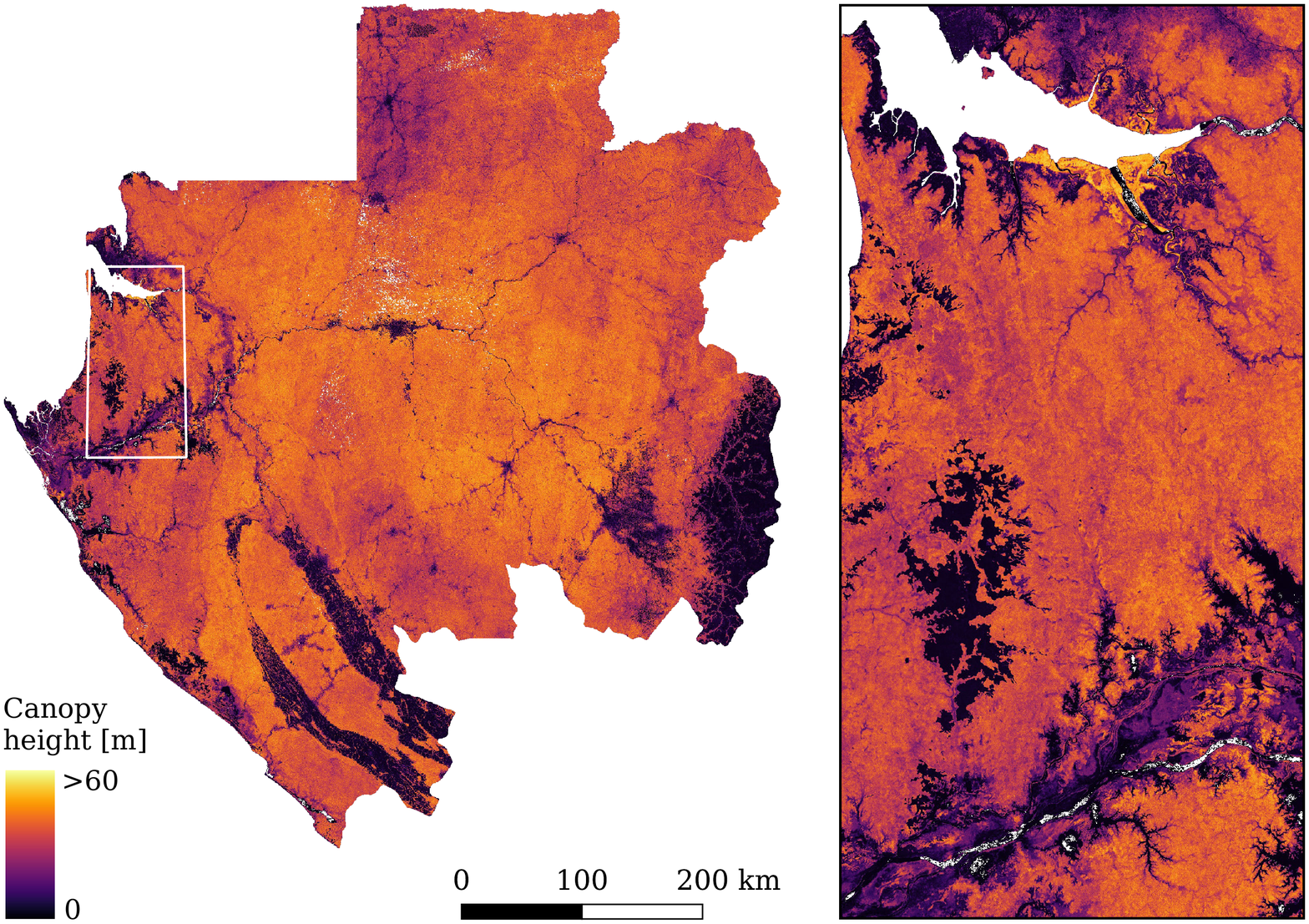}
  		\caption{Our map for 2017 (10~m GSD)}
    \end{subfigure}
    \\
    \vspace{0.5cm}
	\begin{subfigure}[]{0.49\textwidth}
		\centering
  		\includegraphics[width=\textwidth]{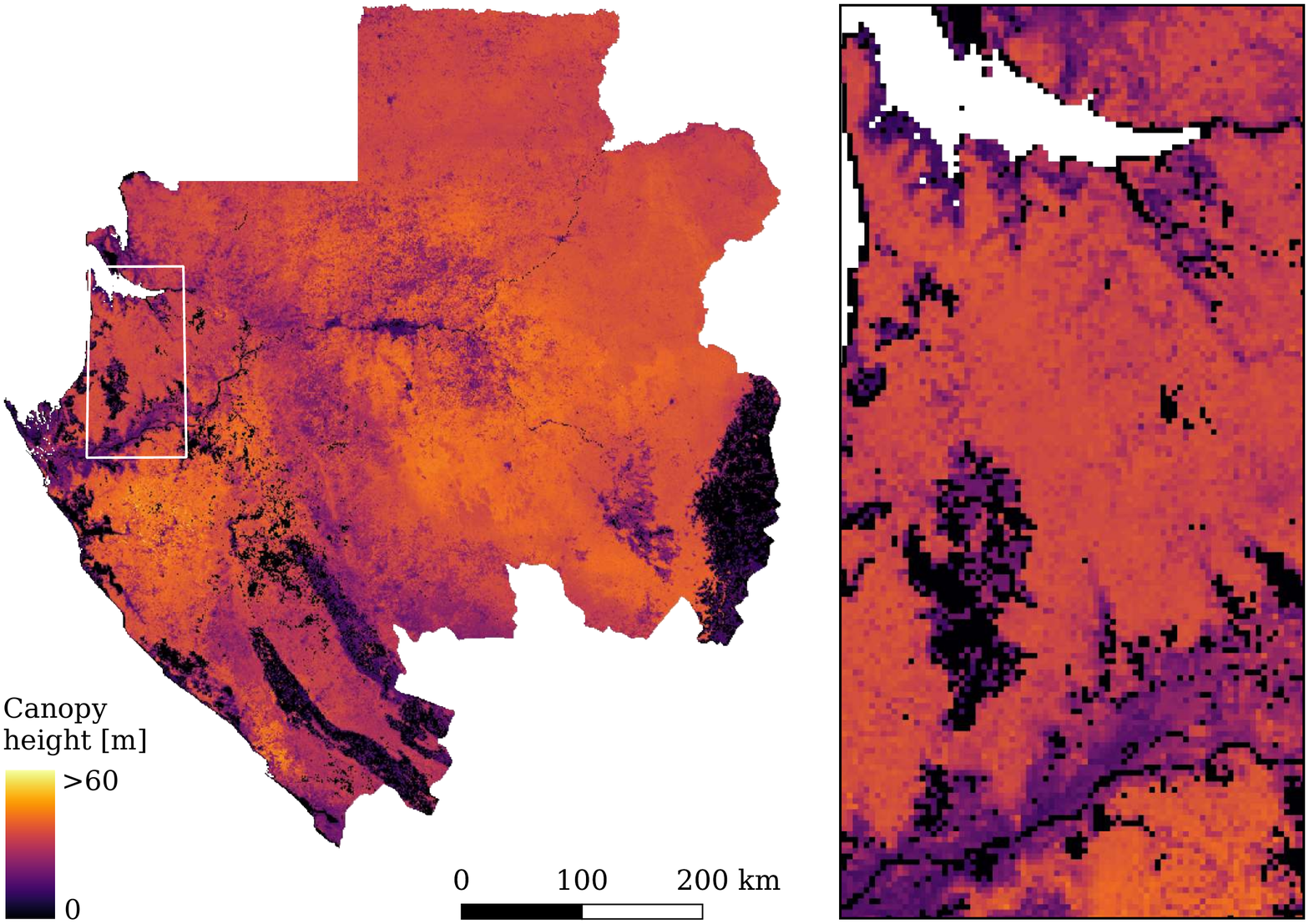}
  		\caption{NASA map for 2005 (1~km GSD)}
    \end{subfigure}
    %
    %
    	\begin{subfigure}[]{0.49\textwidth}
		\centering
  		\includegraphics[width=\textwidth]{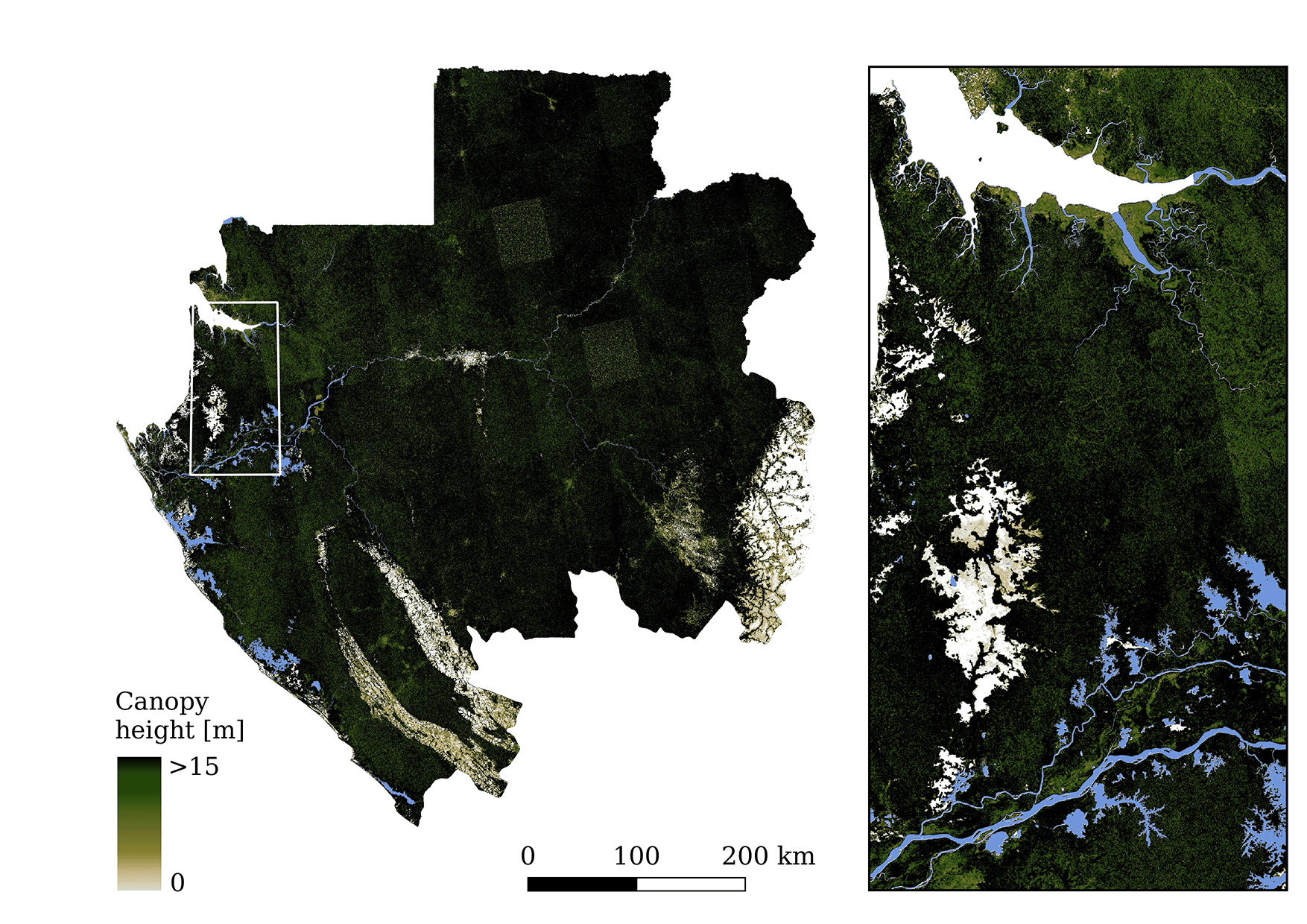}
  		\caption{WHRC map (30~m GSD)}
    \end{subfigure}

\caption{Our country wide vegetation height map of Gabon (top row)
  compared to existing maps from NASA \citep[][bottom
    left]{simard2011mapping} and WHRC \citep[][bottom
    right]{woodshole}. The latter is only available as RGB colour
  image, hence we depict it with the original colour map. Note that
  the height saturates at 15~m.}
\label{fig_GAmap}
\end{figure*}

\begin{figure*}[t]
\centering
	\begin{subfigure}[]{0.45\textwidth}
		\centering
  		\includegraphics[width=\textwidth]{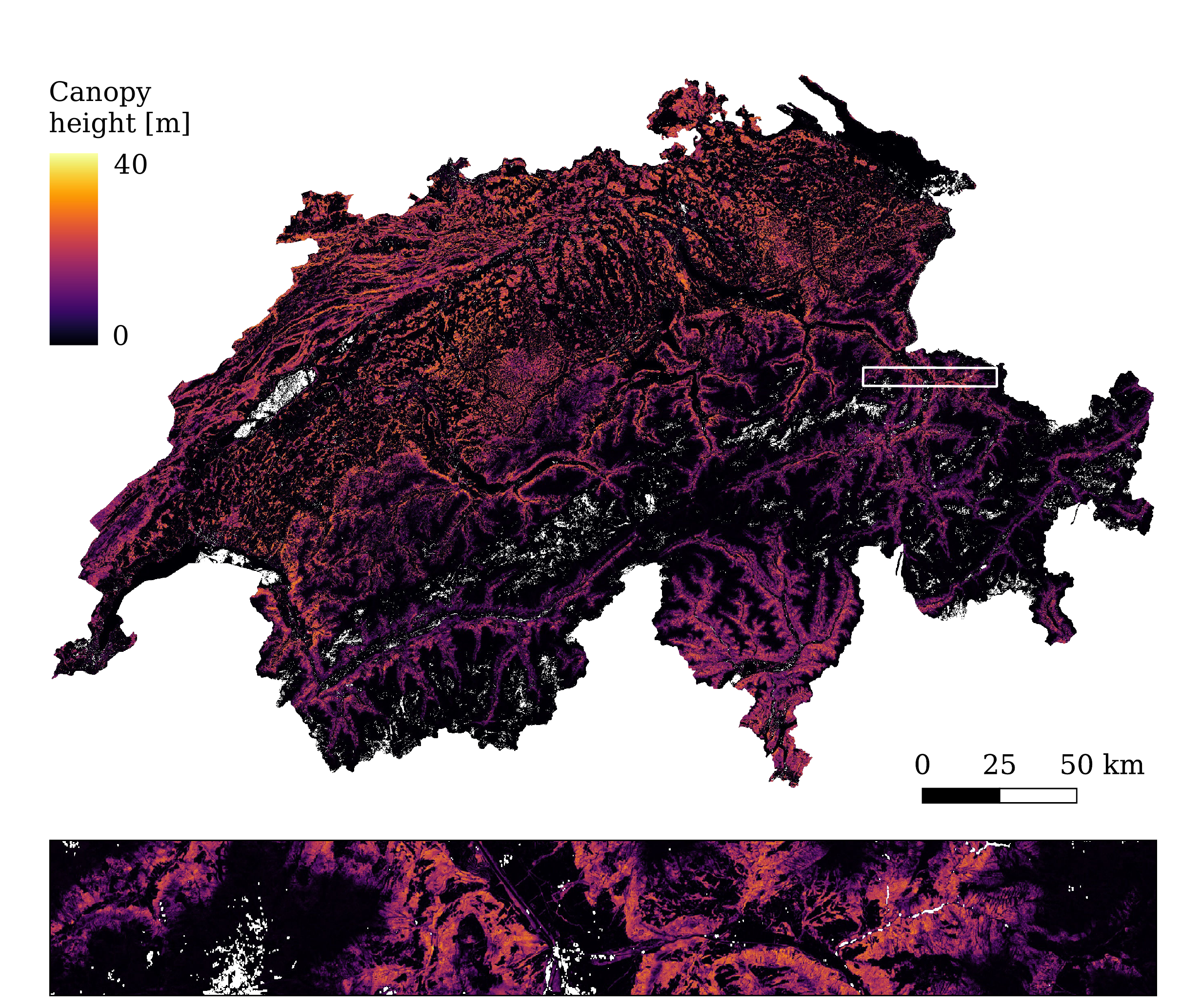}
  		\caption{Our map for 2017 (10~m GSD)}
    \end{subfigure}
    \\
    \vspace{0.5cm}
	\begin{subfigure}[]{0.45\textwidth}
		\centering
  		\includegraphics[width=\textwidth]{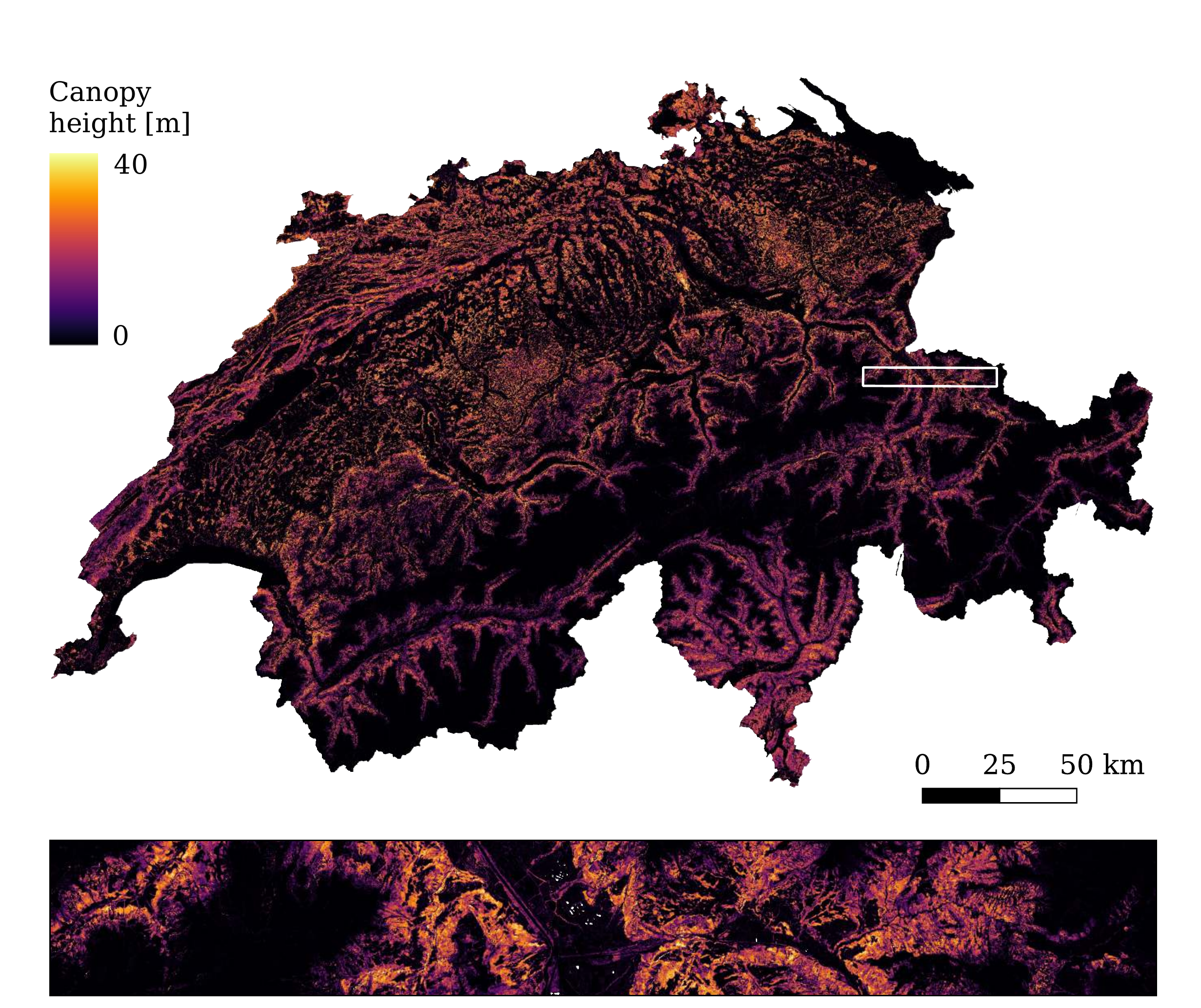}
  		\caption{WSL map 2007-2016 (10~m GSD)}
    \end{subfigure}
    \begin{subfigure}[]{0.45\textwidth}
		\centering
  		\includegraphics[width=\textwidth]{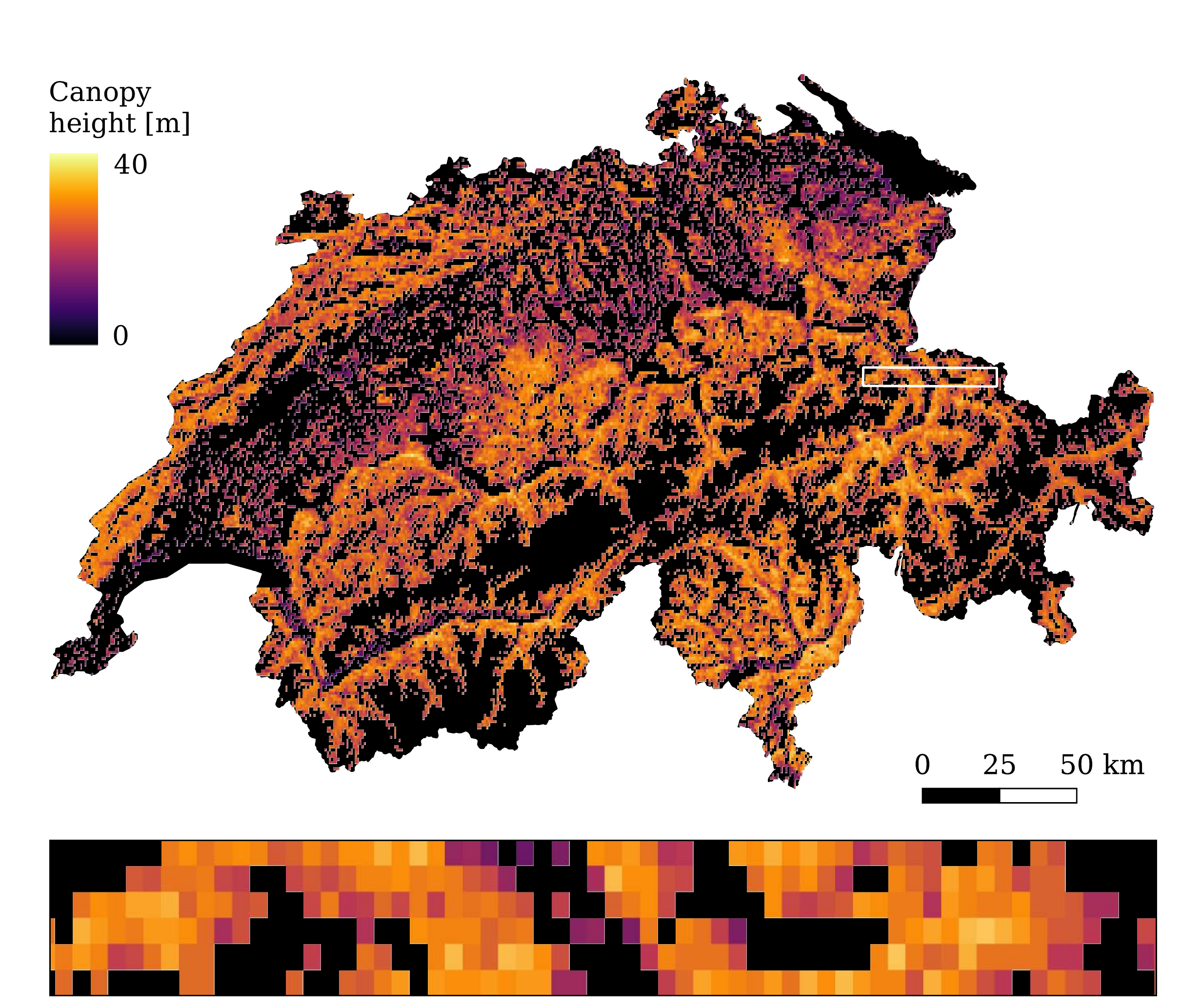}
  		\caption{NASA map for 2005  (1~km GSD)}
    \end{subfigure}

\caption{Our country wide vegetation height map of Switzerland (top)
  compared to existing maps from WSL
  \citep[][bottom left]{ginzler2015countrywide} and NASA
  \citep[][bottom right]{simard2011mapping}.}
\label{fig_CHmap}
\end{figure*}

\begin{figure*}[tb]
	\centering
	\includegraphics[width=0.7\textwidth]{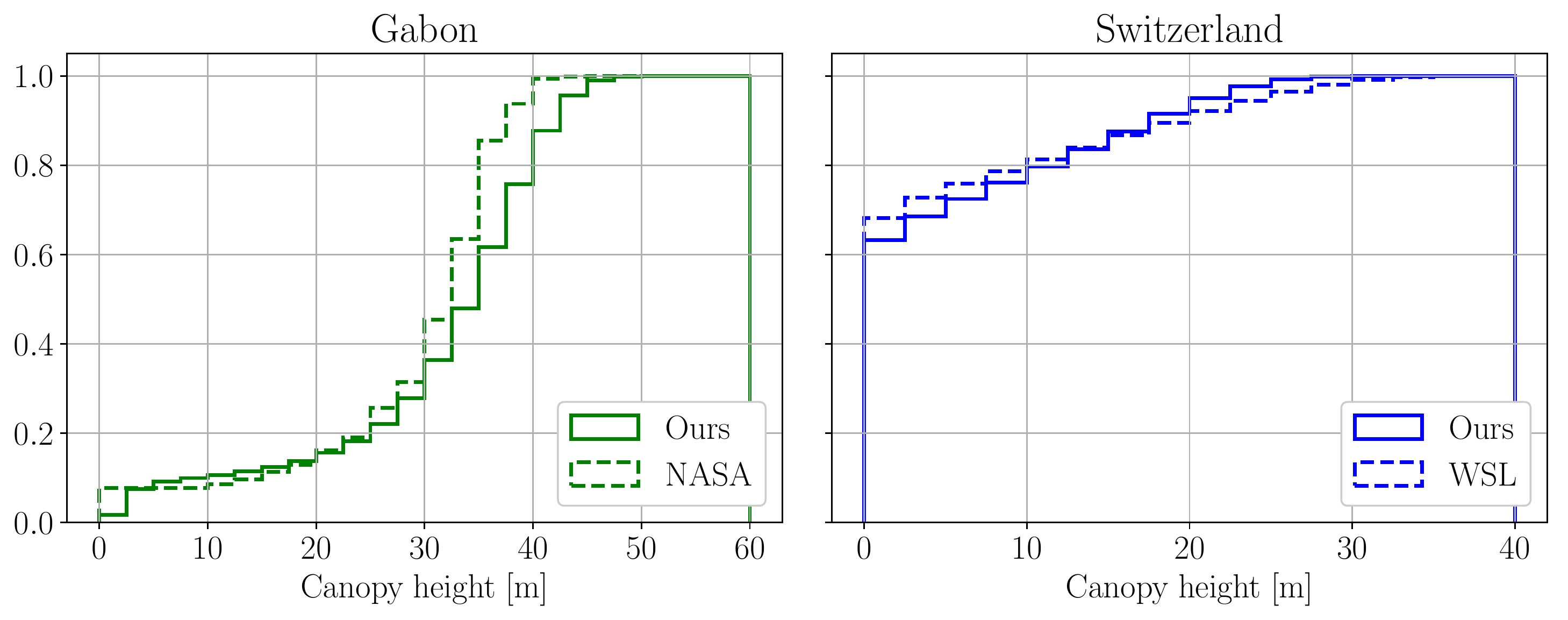}
	\caption{Cumulative distribution of country-wide canopy height
          predictions.}
	\label{fig_countryDistribution}
\end{figure*}

We have also computed country-wide canopy height maps at 10~m
resolution for Gabon (Figure~\ref{fig_GAmap}) and Switzerland (Figure~\ref{fig_CHmap}). Sentinel-2 images are organised
in \linebreak100$\times$100~km$^2$ tiles, such that 56,984 tiles cover the globe.

Hence, 47 tiles are needed to cover Gabon, 13 are needed for Switzerland. To
ensure cloud-free coverage of (almost) the entire country, we
automatically pick, for every tile, the 10 dates within 2017
(May--September for Switzerland) that have the lowest cloud coverage,
predict canopy heights for all cloud-free pixels, and take the median
over the cloud-free predictions at each pixel.

We qualitatively compare the resulting vegetation height maps to
existing maps from NASA \citep[][both locations]{simard2011mapping},
WHRC \citep[][Gabon; map available only for the tropics]{woodshole},
and WSL \citep[][map produced locally for
  Switzerland]{ginzler2015countrywide}. Temporal offsets are
inevitable: our maps represent the state in 2017. The others mix
inputs over longer time periods, due to the recording schedules of the
underlying image and/or reference data.
Both, the NASA map and the WHRC map are fitted to ICESat GLAS LiDAR
heights. While \cite{simard2011mapping} used global high-level
products such as tree cover, elevation, and climatology to interpolate
the LiDAR samples to a global 1~km grid, \cite{woodshole} fit a
regression from ALOS PALSAR to the LiDAR heights, with 30~m GSD.
For the WSL map \citep{ginzler2015countrywide} DSMs were generated
with airborne photogrammetry and reduced to vegetation heights with an
existing country-wide bare-earth DTM. The original data has nominal
GSD 1~m, we downsampled it to 10~m for a direct comparison.

For Gabon, we find that the large-scale structures are in good
agreement between all three maps. The NASA map tends to miss small
structures due to its coarser resolution. While this is not an issue
for large-scale studies, e.g., of the global climate; detailed
information is important for local decisions, such as the protection
of high forest near settlements, as recommended by the High Carbon Stock (HCS) approach.
Moreover, in the NASA map very high regions, like the Mangrove forest
in Pongara National Park near the top of the zoomed region, are
missed. Also the somewhat lower heights around central Gabon's Lope
National Park appear implausible, we speculate that the very frequent
cloud coverage in that region may have impacted some of the products
on which the map is based.
In that region our map still has a few missing pixels, which were
cloud-covered in all 10 selected images. Further images for that tile
need to be processed to fill the gaps.
In the north of the country, around the city of Oyem, our model
predicts visibly lower heights than NASA. At this point we do not have
ground truth to determine which prediction is more correct.

The WHRC map appears to be very accurate in regions of low or missing
vegetation, as expected for a RADAR-based estimate. On the contrary,
strong under-estimates are observed in wetlands such as Pongara. In
general, the low 15~m cut-off is a serious limitation in Gabon, where
practically all forest is higher than 30~m, such that the map largely
degenerates into a binary forest layer -- the overwhelming majority of
apparent height structures on closed canopies are in fact noise. 
According to our prediction (see cumulative distribution in Figure
\ref{fig_countryDistribution}) only $\approx$10\% of Gabon's area has
vegetation height $<$15~m. More than 60\% of Gabon%
\footnote{Gabon's wealth of high forests is indeed
  impressive. Assuming a single tree per 10~m pixel, stacking all those
  trees would reach a height $>$220$\times$ the Earth-Moon distance.} %
is covered with forest higher than 30~m, and $>$20\% are $>$40~m. While
the NASA map indicates an average canopy height of 29.6~m, our
prediction yields an average of 32.1~m.

For Switzerland, the inter-comparison results look rather
different. The WSL map and ours show good agreement. Our result looks
a bit smoother, which is likely a result of the implicit smoothing
when using large receptive fields. Moreover, like in the quantitative
evaluation, our model systematically tends to predict slightly lower
values for very high vegetation.
The NASA map appears to strongly over-estimate both the vegetated area
and the vegetation height throughout Switzerland. The reason for this
behaviour is unclear, we speculate that it may partly be due to some
uncompensated influence of the steep, mountainous topography on the
underlying input products.

In both of our large-scale maps some boundary artifacts remain at the
borders of the satellite passes. Bigger overlaps and/or cross-fading
between adjacent images may mitigate this effect. Furthermore, it may
be beneficial to perform additional radiometric adjustment at the
borders of the input tiles.

Processing a 10,000~km$^2$ (bottom-of-atmosphere\linebreak[4] reflectance) tile with
our model takes $\approx$35 minutes on one GPU. In total we used 274
GPU-hours for Gabon (470 tiles) and 76 GPU-hours for Switzerland (130
tiles).
It has been reported that 94,093 Sentinel-2 tiles are needed to create
a cloud-free composite of the global land
masses~\citep{kempeneers2017optimizing}.
From this we can estimate a processing time of a bit over 6 GPU-years
for a world-wide map. While this may seem a lot, it is equivalent to
2.3 days on a computing cluster with 1000 GPUs -- which nowadays is
feasible without major problems. 
%
  

\section{Conclusion}
Our proposed data-driven approach allows one to map vegetation height
at 10~m resolution. We show that the regression from few Sentinel-2
images achieves low error in the tropics as well as in central Europe,
and that our method is suitable for country-scale canopy height mapping
in terms of generalisation and computation time.
Our CNN-based learning engine, which is able to exploit spatial
context and texture features, can predict a \add{high-resolution}\rem{dense} vegetation height
map from a single cloud-free image with good accuracy. Based on these
findings, we are convinced that global-scale vegetation height mapping
at an unprecedented 10~m resolution can be done operationally,
especially in combination with new data sources like the GEDI
mission~\citep{gedi}.
That resolution would enable a new set of applications requiring
localized vegetation structure, e.g., targeted forest protection in
the context of the High Carbon Stock approach, or fine-grained
biodiversity studies relating vegetation heterogeneity to species
richness.
Ultimately, retrieval from Sentinel-2 could in some cases reduce the
need for expensive LiDAR flight campaigns.
Besides the high spatial resolution, Sentinel-2 provides a new image
every 5 days, which enables frequent updates, and helps to obtain good
coverage even in regions with frequent clouds.
That capability, in turn, opens up the possibility to continuously
monitor forest degradation and loss around settlements
and agricultural lands.

A limitation of our current approach, to be addressed in future work,
is its rudimentary use of multi-temporal information. Spectral time
series could potentially correlate even better with canopy height.
However, we found that satisfactory performance can be reached without
relying on long cloud-free time series.

For us, the most important question is: how to extend our method to
global coverage?
\add{While Sentinel-2 data is globally available, the current models
  are limited by the available training data and will likely not
  generalise well to unseen regions of the world.}
A viable approach, albeit logistically challenging, could be to
collect diverse training data for different vegetation types from
around the world -- GEDI can be seen as a means to that end\add{, even
  though its footprint of 25~m on the ground may not support a 10~m
  output resolution}. \add{Although it is not possible to quantify in
  advance the amount of reference data required to build a globally
  applicable model, GEDI will likely be enough.}
\add{Besides more diverse training data, the fusion of Sentinel-2 with
  other map layers, e.g., elevation as additional inputs to the CNN
  could improve generalisation. Note, however, that several important
  drivers of tree growth, like temperature or precipitation, are not
  available at the desired, high resolution.}
A scientifically exciting, but perhaps more risky approach would be to
attempt unsupervised domain adaptation \add{to new geographic
  regions}, based on the statistics of unlabeled Sentinel-2 data
\add{that are available worldwide}. \rem{from different regions.}

\newpage
\section{Acknowledgement}
We thank Christian Ginzler from WSL for sharing the reference data for
Switzerland. We greatly appreciate the open data policies of the LVIS
project and the ESA Copernicus program.
The project received funding from Barry Callebaut Sourcing AG, as a
part of a Research Project Agreement.



\bibliography{treeheight-sentinel}


\end{document}